\documentclass[12pt]{article}
\pdfoutput=1
\usepackage{amssymb,amsmath}
\usepackage{graphicx}
\usepackage{color}
\usepackage[colorlinks=true
,urlcolor=blue
,citecolor=blue
,linkcolor=blue
,pagecolor=blue
,linktocpage=true
,pdfproducer=medialab
]{hyperref}
\usepackage[a4paper]{geometry}

\usepackage{placeins}
\usepackage{slashed}
\usepackage{mathtools}
\usepackage[FIGTOPCAP]{subfigure}

\makeatletter \renewcommand{\@dotsep}{10000} \makeatother

\mathchardef\mhyphen="2D

\newcommand{\beq}{\begin{equation}}
\newcommand{\eeq}{\end{equation}}
\newcommand{\bea}{\begin{eqnarray}}
\newcommand{\eea}{\end{eqnarray}}

\usepackage[nodisplayskipstretch]{setspace}

\begin{document}

\begin{titlepage}
\pagestyle{empty}

\vspace*{0.2in}
\begin{center}
{\Large \bf  Gluino Search with Stop and Top in \\ Nonuniversal Gaugino Mass Models at \\ LHC and Future Colliders}\\
\vspace{1cm}
{\bf  Zafer Alt\i n$^a$\footnote{E-mail: 501407009@ogr.uludag.edu.tr},
Ali  \c{C}i\c{c}i$^a$\footnote{E-mail: 501507007@ogr.uludag.edu.tr},
Zerrin K\i rca$^a$\footnote{E-mail: zkirca@uludag.edu.tr},
Qaisar Shafi$^b$\footnote{shafi@bartol.udel.edu} 
 and
Cem Salih $\ddot{\rm U}$n$^a$\footnote{E-mail: cemsalihun@uludag.edu.tr}}
\vspace{0.5cm}

{\it $^a$Department of Physics, Bursa Uluda\~{g} University, TR16059 Bursa, Turkey} \\
{\it $^b$ Bartol Research Institute, Department of Physics and Astronomy, University of Delaware, Newark, DE 19716, USA}

\end{center}

\vspace{0.5cm}
\begin{abstract}
\noindent

We discuss the gluino mass in the CMSSM and {Nonuniversal Gaugino Mass Models (NUGM)} frameworks in light of the results from the current LHC and {Dark Matter} experiments. Assuming negative results from the current and near future LHC experiments, we probe the gluino mass scales by considering its decay modes into stop and top quarks, $\tilde{g}\rightarrow \tilde{t}_{1}t$ and $\tilde{g}\rightarrow \bar{t}t\tilde{\chi}_{1}^{0}$, {where $\tilde{t}_{1}t$ represents both $\tilde{t}_{1}\bar{t}$ and $\tilde{t}_{1}^{*}t$.} The region with $m_{\tilde{g}} \lesssim 2$ TeV  is excluded up to $68\%$ {CL} in the CMSSM {if} the gluino {decays} into a stop and top quark, while the $95\%$ {CL exclusion requires} $m_{\tilde{g}}\gtrsim 1.9$ TeV. Considering {an} error of about $10\%$ in {calculations} of the SUSY mass spectrum, such exclusion bounds on the gluino mass more or less overlap with the current LHC results. The {decay mode} $\tilde{g}\rightarrow \bar{t}t\tilde{\chi}_{1}^{0}$ may take over {if} $\tilde{g}\rightarrow \tilde{t}_{1}t$ is not allowed. One can probe the gluino mass {in this case} up to about 1.5 TeV with $68\%$ CL in the CMSSM, and about 1.4 TeV with $95\%$ CL. {Imposing the Dark Matter constraints yields a lower bound on} the gluino and stop {masses of} about 3.2 TeV from below, which is beyond the reach of the current LHC experiments. {A similar} analyses in the NUGM {framework} yield exclusion curves for the gluino mass $m_{\tilde{g}}\gtrsim 2.1$ TeV {at 14 TeV} for both decay modes of the gluino under consideration. We also show that {increasing} the center of mass energy to 27 TeV can probe the gluino mass up to about 3 TeV {through} its decay {mode} into {stop and top quark}. The {Dark Matter} constraints are not very severe in the framework of NUGM, and they allow solutions with $m_{\tilde{g}},m_{\tilde{t}} \gtrsim 1$ TeV. {In addition, with} NUGM the LSP neutralino can {coannihilate with} gluino and/or stop {for} $m_{\tilde{g}},m_{\tilde{t}}\approx m_{\tilde{\chi}_{1}^{0}} \in [0.9-1.5]$ TeV. {{With the 100 TeV FCC collider one can probe the gluino masses up to about 6 TeV with $36.1~fb^{-1}$ integrated luminosity.} We also find that the {decay} $\tilde{g}\rightarrow \tilde{t}t$ can indirectly probe {the} {stop mass} up to about 4 TeV.}

\end{abstract}

\end{titlepage}

\setcounter{footnote}{0}

\section{Introduction}
\label{ch:introduction}


{The low scale implications of Supersymmetric (SUSY) SO(10) grand unified theories (GUTs) such as third family Yukawa unification and sparticle spectroscopy in SO(10) \cite{big-422,bigger-422} and SU(5) \cite{Chattopadhyay:2001mj} have occupied a fair amount of attention in recent years. The current LHC results exclude a gluino lighter than about 2.1 TeV \cite{Aaboud:2017vwy}, unless it happens to be the Next to Lightest Supersymmetric Particle (NLSP), in which case the bound on its mass can go down about {800} GeV. The bound on the stop mass varies depending on its decay modes. Thus, $m_{\tilde{t}}\gtrsim 1200$ GeV if} $\tilde{t}\rightarrow t\tilde{\chi}_{1}^{0}$ and $m_{\tilde{t}}\gtrsim 1100$ GeV {for} $\tilde{t}\rightarrow b\tilde{\chi}_{1}^{\pm}$ \cite{Vami:2019slp}. These analyses have been performed mostly for low scale SUSY, where the SUSY spectrum can be adjusted such that the particles participating in the analyzed decay modes are light enough, while the rest of the spectrum is {appropriately} heavy and cannot interfere in the stop and gluino decays. Since {a generic supersymmetric model has} more than a hundred free parameters, such {assumptions seem plausible}. Besides, the probability for the relevant decay modes under the analyses (branching fractions, for instance) can be optimized. 

On the other hand, the SUSY GUTs {SO(10) and SU(5)} can significantly reduce the number of free parameters, and some of the low scale {assumptions} may not be possible, when all low scale observables are calculated in terms of a few free parameters. For instance, one of our recent studies has shown that the solutions with $m_{\tilde{t}}\lesssim 500$ GeV can be excluded only within $60\%$ confidence level (CL) for $\tilde{t}\rightarrow t\tilde{\chi}_{1}^{0}$, and {at} $50\%$ CL for $\tilde{t}\rightarrow b\tilde{\chi}_{1}^{\pm}$ \cite{Cici:2016oqr}. {For $m_{\tilde{t}} \gtrsim 500$ GeV} the exclusion is {realized} at the order of a few percent, even though the current results exclude {a stop mass} below about 800 GeV, {if} these decay modes are allowed in the low scale analyses. 

A similar discussion can be {applied to} the gluino. {The current results exclude solutions with $m_{\tilde{g}}\lesssim 2.4$ TeV, when the LSP neutralino is of mass around a TeV, while the mass bound on the gluino is reduced to about 2.2 TeV, as the LSP neutralino mass decreases \cite{Vami:2019slp}}. Even though the sleptons and/or sneutrinos are light enough to take part, the {lower} mass bound on the gluino is still about 2 TeV \cite{Leblanc:2018rfd}. {A} 125 GeV Higgs boson mass constrains the stop masses at {around a} TeV scale from {below}, {but} it is still possible to include the stop in {the} gluino decay {mode} $\tilde{g}\rightarrow \tilde{t}t$. {If} this decay mode is kinematically allowed, the bound on the gluino mass can be lowered to about 1.8 TeV \cite{Aad:2016eki}, {while it has recently been reported by the CMS analyses as $m_{\tilde{g}}\gtrsim 1.6$ TeV \cite{Vami:2019slp}}.


In this study, we consider the {gluino} decay into a stop and top quark, and apply the constraints from low scale analyses. Since the sbottom usually happens {to be} much heavier than stops (see, for instance, the benchmark points represented in several papers in Ref. \cite{bigger-422}), we do not consider processes in which the gluino decays into a sbottom and bottom quarks. Hence, the decay cascades {considered} are the following:

\begin{equation}\hspace{-7.0cm}
 p p \rightarrow \tilde{g}\tilde{g}\xrightarrow{\text{\large $\tilde{g}\rightarrow \tilde{t}t$}}\tilde{t}\tilde{t}^{*}t\bar{t}\xrightarrow{\text{ \large $\tilde{t}\rightarrow t\tilde{\chi}_{1}^{0}$}}t\bar{t}t\bar{t}\tilde{\chi}_{1}^{0}\tilde{\chi}_{1}^{0}~,
\label{eq:case1}
\end{equation}
\begin{equation}
 p p \rightarrow \tilde{g}\tilde{g}\xrightarrow{\text{\large $\tilde{g}\rightarrow \tilde{t}t$}}\tilde{t}\tilde{t}^{*}t\bar{t}\xrightarrow{\text{ \large $\tilde{t}\rightarrow b\tilde{\chi}_{1}^{\pm}$}}b\bar{b}t\bar{t}\tilde{\chi}_{1}^{\pm}\tilde{\chi}_{1}^{\pm}\xrightarrow{\text{ \large $\tilde{\chi}_{1}^{\pm}\rightarrow W^{\pm}\tilde{\chi}_{1}^{0}$}}b\bar{b}t\bar{t}W^{\pm}W^{\pm}\tilde{\chi}_{1}^{0}\tilde{\chi}_{1}^{0}~.
\label{eq:case2}
\end{equation}

The process given in Eq.(\ref{eq:case1}) {will} be called Signal1, while Signal2 refers to the process given in Eq.(\ref{eq:case2}). If the gluino decay mode involving a stop and top quarks is not likely, then one can consider another mode {denoted} as Signal3, in which the gluino decays directly into a LSP neutralino along with a pair of top quarks:

\begin{equation}
p p \rightarrow \tilde{g}\tilde{g}\xrightarrow{\text{\large $\tilde{g}\rightarrow \bar{t}t\tilde{\chi}_{1}^{0}$}}\bar{t}t\bar{t}t\tilde{\chi}_{1}^{0}\tilde{\chi}_{1}^{0}~.
\label{eq:case3}
\end{equation}

As in the low scale analyses, {we start} with the gluino pair production and the next step of the decay cascade includes one pair of stops and one pair of top quarks. After this step, the analyses may resemble the stop quark analyses, {where the} stop can decay {through either of the two processes}, $\tilde{t}\rightarrow t\tilde{\chi}_{1}^{0}$ and $\tilde{t}\rightarrow b\tilde{\chi}_{1}^{\pm}$. The strongest {constraint} in the latter case arises {if} the chargino is allowed to decay into a LSP neutralino {and} a $W^{\pm}$ boson. Note that it is also possible that the stop can decay into a charm quark and LSP neutralino; however, the signal is soft in this case, and the {constraint} is not very severe ($m_{\tilde{t}}\gtrsim 230$ GeV \cite{TheATLAScollaboration:2013aia}). Thus, we do not consider {this latter case in the following discussion}. The processes given above continue with the decays such as $t\rightarrow b W^{\pm}$ and $W^{\pm}\rightarrow l\nu$. 

The third {stage} (fourth in the process given in Eq.(\ref{eq:case2})) in the decay cascade of the signal includes two pairs of the third family quarks and two LSP neutralinos, if the stop decays into a top quark and LSP neutralino. The second option involves two W-bosons, {and} pairs of top and bottom quarks along with two neutralinos. The relevant background processes can be listed  as $t\bar{t}$, single top, $t\bar{t}W$, $t\bar{t}Z$, $t\bar{t}t\bar{t}$, $t\bar{t}h$, $WW$, $WZ$, $ZZ$ and $W/Z+{\rm jets}$ \cite{Aad:2016eki}. Even though the top quark pair production dominates in the background processes, the signal is expected to have much more missing energy, and the events from the top pair production background can be suppressed by applying a cut on the missing energy as $\slashed{E}_{T} \gtrsim 300$ GeV \cite{Cici:2016oqr}. In this case, production of {two pairs} top quarks can be considered the main {background} in its final state.

{Following} \cite{Cici:2016oqr}, as a benchmark study, our aim {here} is {to explore} the gluino searches {in the framework of}  SUSY GUTs via the decay mode $\tilde{g}\rightarrow \tilde{t}t$. {The} analyses of \cite{Cici:2016oqr} will be generalized by employing the constraints and calculations to the whole data generated within the SUSY GUT framework. The paper is organized as follows. Section \ref{sec:scan} summarizes the scanning procedure and the experimental constraints employed in our analyses. {In Section \ref{sec:CMSSM} we briefly} present the results and {constraints} on the gluino mass in CMSSM. Since the universal boundary conditions of CMSSM lead to a linear correlation between the gluino and LSP neutralino masses, we generalize our discussion {in Section} \ref{sec:NUGM} by allowing non-universal gaugino masses (NUGM) at the GUT scale. {We describe the constraints on} the gluino mass {from} the current LHC experiments, as well as {from} the High Energy LHC (HE-LHC at 27 TeV) and Future Circular Collider (FCC at 100 TeV). {Our conclusions are summarized} in Section \ref{sec:conc}.

\section{Scanning Procedure and Experimental Constraints}
\label{sec:scan}

We have employed SPheno 4.0.3 package \cite{Porod:2003um, Porod2} {generated} with SARAH 4.13.0 \cite{Staub:2008uz,Staub2}. In this package, the weak scale values of the gauge and Yukawa couplings in MSSM are evolved to the unification scale $M_{{\rm GUT}}$ via the renormalization group equations (RGEs). $M_{{\rm GUT}}$ is determined by the requirement {of unification of the gauge couplings} through their RGE evolutions. Note that we do not strictly enforce the unification condition $g_1 = g_2 = g_3$ at $M_{{\rm GUT}}$ since a few percent deviation from the unification can be assigned to unknown GUT-scale threshold corrections \cite{Hisano:1992jj,GUTth}. With the boundary conditions given at $M_{{\rm GUT}}$, all the soft supersymmetry breaking (SSB) parameters along with the gauge and Yukawa couplings are evolved back to the weak scale.

We have performed random scans over the parameter space of CMSSM and NUGM {models} as follows:

\begin{equation}
\centering
\setstretch{1.5}
\scalebox{0.80}{$
\begin{array}{ccc|ccc}
\multicolumn{3}{c|}{{\rm CMSSM}} & \multicolumn{3}{c}{{\rm NUGM}}  \\ \hline
0 \leq & m_{0} & \leq 5~{\rm (TeV)} & 0 \leq & m_{0} & \leq 5~{\rm (TeV)}  \\
0 \leq & M_{1/2} & \leq 5~{\rm (TeV)} &0 \leq & M_{1}, M_{2}, M_{3} & \leq 5~{\rm (TeV)}  \\
1.2 \leq & \tan\beta & \leq 50 & 1.2 \leq & \tan\beta & \leq 50 \\
-3 \leq & A_{0}/m_{0} & \leq 3 & -3 \leq & A_{0}/m_{0} & \leq 3 ~,
\end{array}$}
\label{paramSP}
\end{equation}
where $m_0$ is the universal SSB mass term for the matter scalars and Higgs fields. $M_{1}$, $M_{2}$ and $M_{3}$ are the SSB mass terms for the gauginos associated with the  $U(1)_{Y}$, $SU(2)_{L}$ and $SU(3)_{C}$ symmetry groups respectively. While the SSB gaugino {masses} are independent terms in NUGM, {in CMSSM they satisfy} $M_{1}=M_{2}=M_{3}=M_{1/2}$ at $M_{{\rm GUT}}$. $A_0$ is {the} SSB trilinear coupling, and $\tan\beta$ is ratio of VEVs of the MSSM Higgs doublets. In addition to these parameters, 
the radiative electroweak symmetry breaking (REWSB) conditions determine the value of {the MSSM} $\mu-term$ but not its {sign, which we assume to be} positive in our scans. Finally, we have used {the} central value of top quark mass, $m_t=173.3$ GeV \cite{Group:2009ad}. Note that the sparticle spectrum is not too sensitive {for} one or two sigma variation in the top quark mass \cite{Gogoladze:2011db}, but it can shift the Higgs boson mass by  1-2 GeV \cite{Gogoladze:2011aa,Ajaib:2013zha}.

The REWSB condition provides a strict theoretical constraint \cite{Ibanez:Ross,REWSB2,REWSB3,REWSB4,REWSB5} over the fundamental parameter space given in Eq.(\ref{paramSP}). Another important constraint comes from the relic abundance of charged supersymmetric particles \cite{Nakamura:2010zzi}. This constraint excludes regions which yield charged particles such as stop and stau {as} the lightest supersymmetric particle (LSP). In this context, we accept only the solutions which satisfy the REWSB condition and yield neutralino LSP. {In this case}, it is also {appropriate} that the LSP {becomes a suitable dark matter candidate}. {The thermal} relic abundance of LSP should, {of course}, be consistent with the current results from the WMAP \cite{Hinshaw:2012aka} and Planck \cite{Akrami:2018vks} satellites. However, even if a solution does not satisfy the dark matter observations, it can still survive in conjunction with other form(s) of the  dark matter \cite{Baer:2012by}. {We} mostly focus on the LHC allowed solutions, {but} we also discuss the DM implications in our results.

In scanning the parameter space we use {an} interface, which employs {the} Metropolis-Hasting algorithm described in  \cite{Belanger:2009ti,SekmenMH}. After generating the low scale data with SPheno, all outputs are transferred to {MicrOmegas} \cite{Belanger:2006is} for calculations of the relic abundance of the LSP neutralino as a candidate for DM. At the final step, we transfer the same output files to MadGraph \cite{Alwall:2011uj} to calculate the cross-sections of the signal and relevant background processes. We should note that the following approximation has been used in the cross-section calculations:

\begin{equation}\hspace{-4.4cm}
\sigma({\rm Signal \ref{eq:case1}})\approx \sigma(pp\rightarrow \tilde{g}\tilde{g})\times {\rm BR}(\tilde{g}\rightarrow \tilde{t}_{1}t)^{2}\times{\rm BR}(\tilde{t}_{1}\rightarrow t\tilde{\chi}_{1}^{0})^{2}~,
\label{eq:cross1}
\end{equation}
\begin{equation}
\sigma({\rm Signal \ref{eq:case2}})\approx \sigma(pp\rightarrow \tilde{g}\tilde{g})\times {\rm BR}(\tilde{g}\rightarrow \tilde{t}_{1}t)^{2}\times{\rm BR}(\tilde{t}_{1}\rightarrow b\tilde{\chi}_{1}^{\pm})^{2}\times{\rm BR}(\tilde{\chi}_{1}^{\pm}\rightarrow W^{\pm}\tilde{\chi}_{1}^{0})^{2}~,
\label{eq:cross2}
\end{equation}
\begin{equation}\hspace{-7.6cm}
\sigma({\rm Signal \ref{eq:case3}})\approx \sigma(pp\rightarrow \tilde{g}\tilde{g})\times {\rm BR}(\tilde{g}\rightarrow \bar{t}t)^{2}~.
\label{eq:cross3}
\end{equation}

Even though MadGraph calculates cross-sections by using full matrix elements, comparisons over a control set of data points reveal {an error of $0.7\%$ at most} between the {MadGraph} results and the approximations given above {at.} Thus, we obtain the results for the gluino pair production by MadGraph, and the full cross-section of the signals are calculated by using the approximation in Eqs.(\ref{eq:cross1}, \ref{eq:cross2} and \ref{eq:cross3}). After collecting the {MadGraph} results, we successively apply the mass bounds on all sparticles \cite{Agashe:2014kda} and the constraints from the rare B-decays ($B_s \rightarrow \mu^+ \mu^-$ \cite{Aaij:2012nna}, $B_s \rightarrow X_s \gamma$ \cite{Amhis:2012bh} and $B_u \rightarrow \tau \nu_\tau $ \cite{Asner:2010qj}). In applying the mass bounds on the supersymmetric particles, we apply the LEP II bound on the gluino \cite{Bisset:1996sb}. The solutions surviving after these constraints {are referred to} as the LHC allowed points, and the DM constraints are applied {on these} solutions. {Regarding} the relic abundance of LSP neutralino, we show bounds both from the WMAP and Planck satellites within $5\sigma$ uncertainty.

The experimental constraints can be listed as follows:

\begin{equation}
\setstretch{2.0}
\begin{array}{l}
123 \leq m_{h} \leq 127~{\rm GeV}\\
 m_{\tilde{g}}  \geq 260~{\rm GeV}\\
 0.8\times 10^{-9} \leq BR (B_s \rightarrow {\mu}^{+} {\mu}^{-}) \leq 6.2 \times 10^{-9}~ (2\sigma) \\
2.9\times 10^{-4} \leq BR (b \rightarrow s {\gamma})\leq 3.87\times 10^{-4}~ (2\sigma)\\
0.15 \leq \dfrac{BR (B_u \rightarrow {\nu}_{\tau} {\tau})_{MSSM}}{BR (B_u \rightarrow {\nu}_{\tau} {\tau})_{SM}}\leq 2.41 ~ (3\sigma) \\
0.0913 \leq \Omega h^{2}({\rm WMAP}) \leq 0.1363 ~(5\sigma) \\
0.114 \leq \Omega h^{2}({\rm Planck}) \leq 0.126 ~(5\sigma)~.
\end{array}
\label{constraints}
\end{equation}
In addition to these constraints, we define the signal significance ($SS$) as 

\begin{equation}
SS=\dfrac{S}{\sqrt{S+B}}~,
\end{equation}
where $S$ and $B$ refer to the event number (cross-section $\times$ Luminosity) of the signal and background respectively. We use the following correspondence to translate $SS$ to the confidence level (CL) \cite{Cranmer:2015nia}:

\begin{equation}
\setstretch{1.5}
\begin{array}{l}
0 \leq SS < 1 \rightarrow {\rm hardly~excluded}~, \\
1 \leq SS < 2 \rightarrow {\rm excluded~up~to~68\%}~, \\
2 \leq SS < 3 \rightarrow {\rm excluded~up~to~95\%}~.
\end{array}
\end{equation}

\section{CMSSM at 14 TeV}
\label{sec:CMSSM}

\begin{figure}[ht!]
\centering
\subfigure{\includegraphics[scale=1.1]{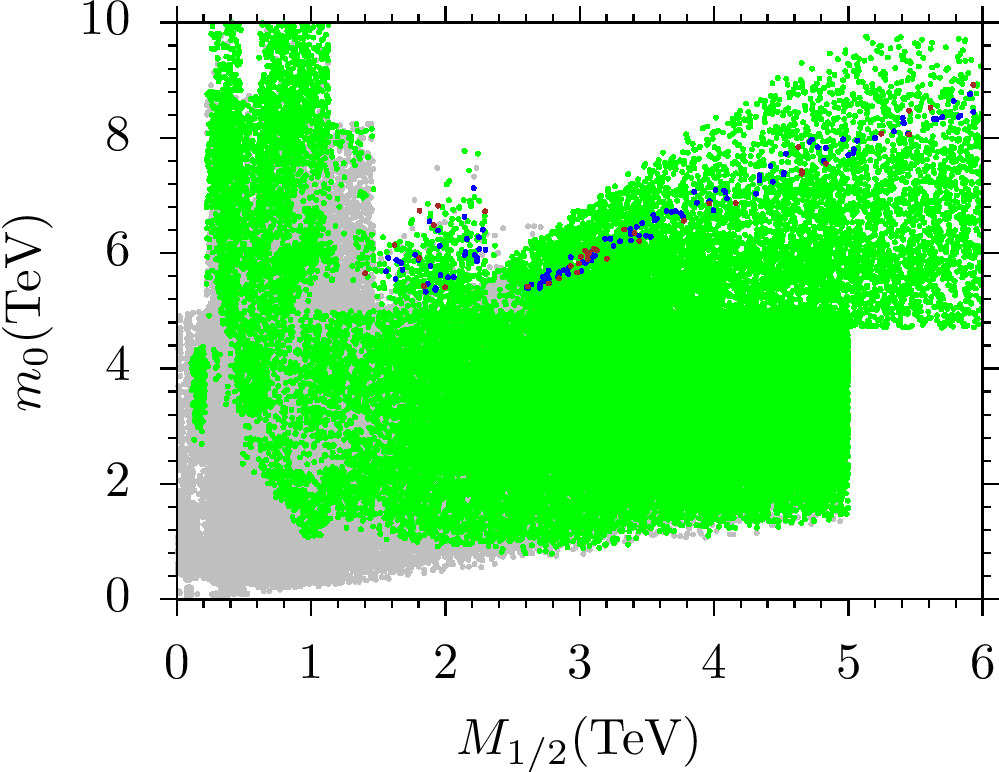}}
\subfigure{\includegraphics[scale=1.1]{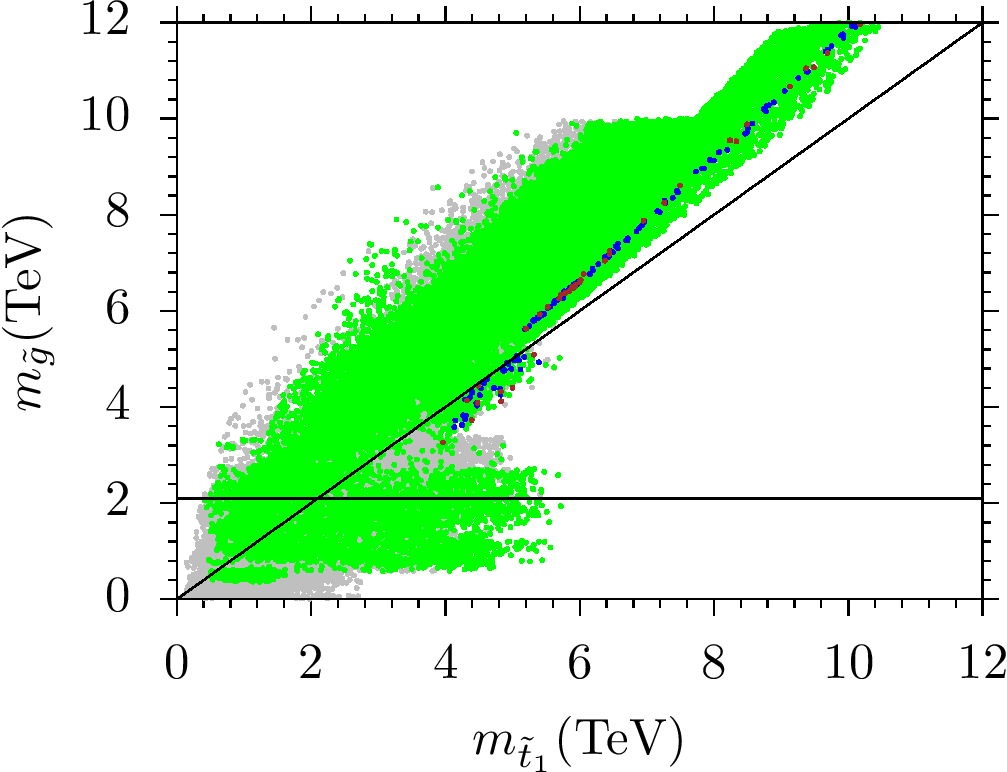}}
\caption{Plots in the $m_{0}-M_{1/2}$ and $m_{\tilde{g}}-m_{\tilde{t}_{1}}$ planes. All points are compatible with REWSB and neutralino LSP {conditions}. Green points are consistent with the current mass bounds and constraints from rare B-meson decays except {for} gluino, on whose mass the LEP II bound is applied. {Blue} points form a subset of green, and they indicate the solutions allowed by the WMAP bound on the relic abundance of neutralino LSP within $5\sigma$; brown points are a subset of blue and they satisfy the Planck bounds on the relic abundance within $5\sigma$. The horizontal line in the right panel indicates the current mass bound on the gluino mass, and the diagonal line {indicates the mass degeneracy between the stop and gluino}.}
\label{fig1}
\end{figure}

\begin{figure}[ht!]
\centering
\subfigure{\includegraphics[scale=1.1]{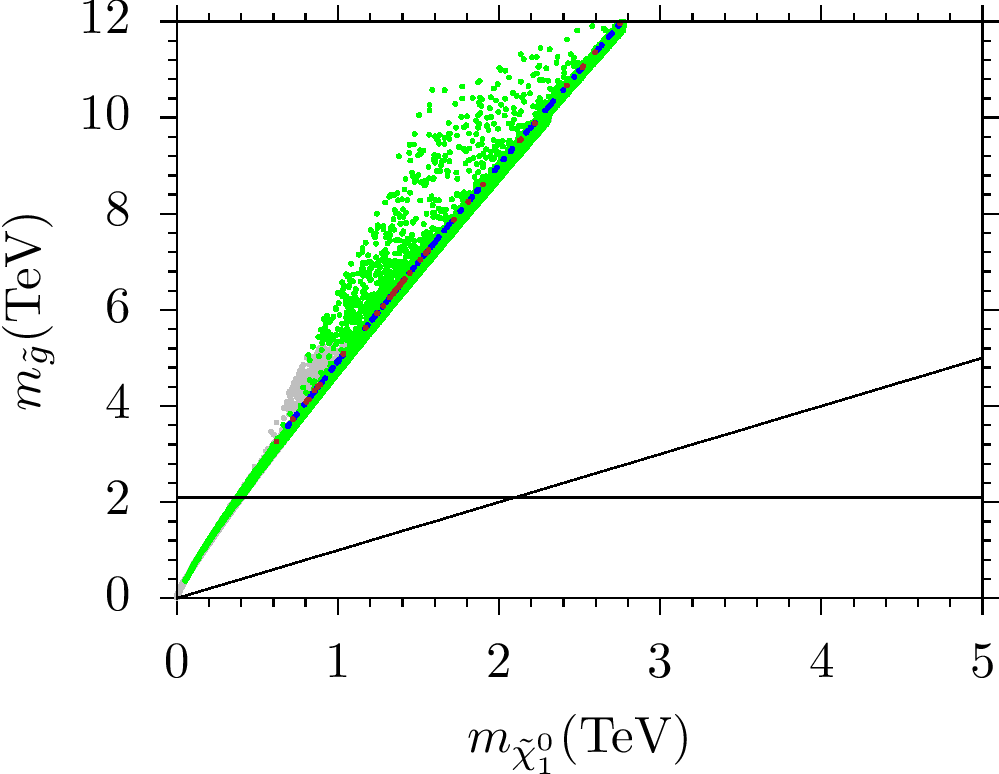}}
\subfigure{\includegraphics[scale=1.1]{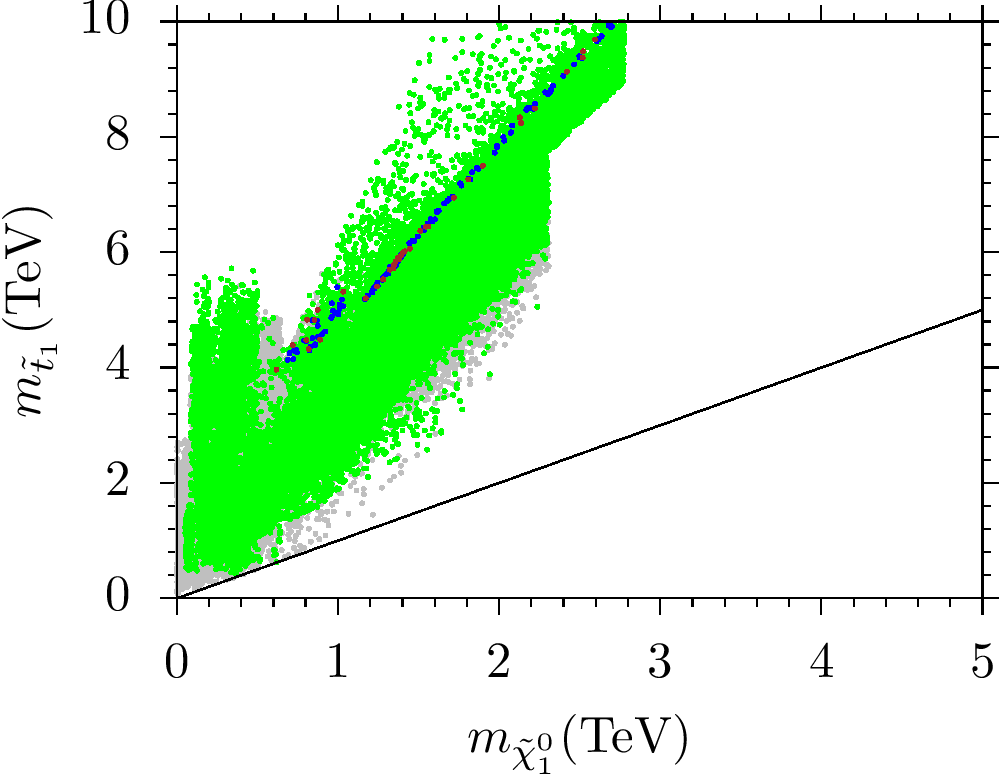}}
\subfigure{\includegraphics[scale=1.1]{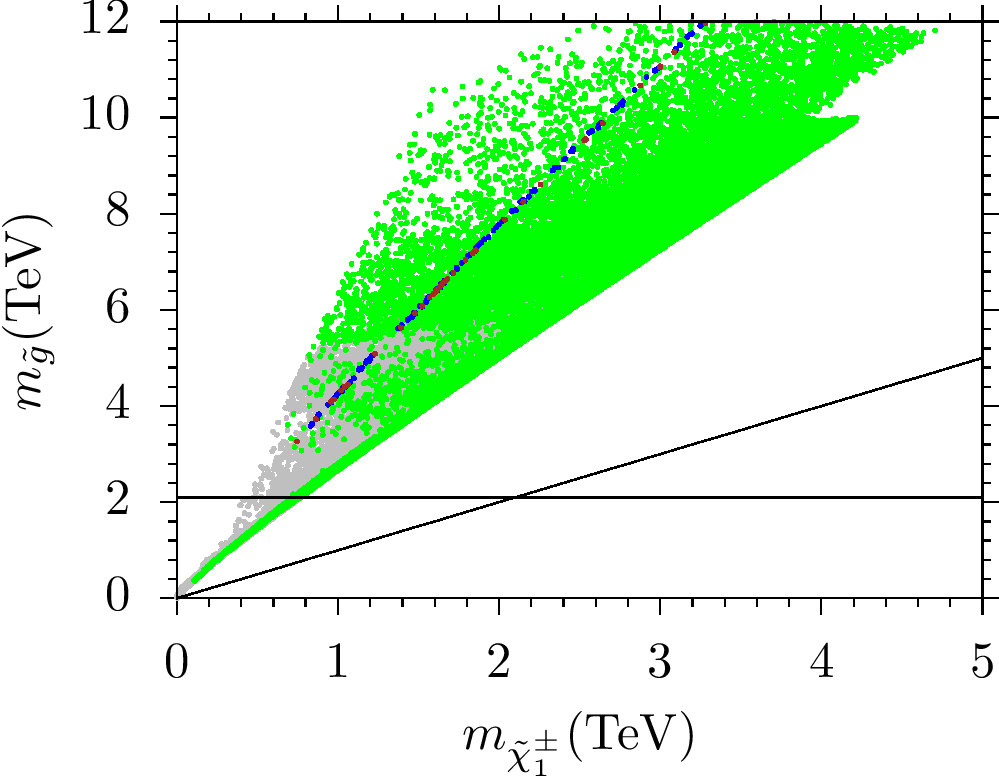}}
\subfigure{\includegraphics[scale=1.1]{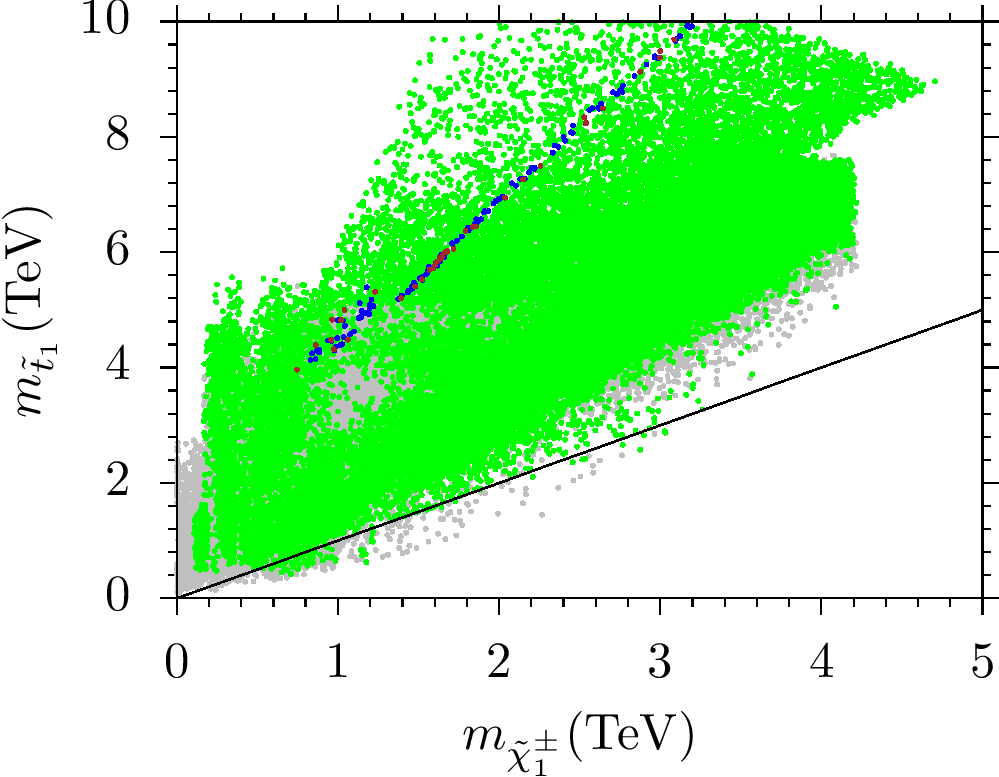}}
\caption{Plots in the $m_{\tilde{g}}-m_{\tilde{\chi}_{1}^{0}}$, ${m_{\tilde{t}_{1}}}-m_{\tilde{\chi}_{1}^{0}}$, $m_{\tilde{g}}-m_{\tilde{\chi}_{1}^{\pm}}$ and ${m_{\tilde{t}_{1}}}-m_{\tilde{\chi}_{1}^{\pm}}$ planes. The color coding is the same as {in} Figure \ref{fig1}. The diagonal lines in the panels show {the degeneracy in mass states of} the plotted particles. The horizontal lines in the {top left and bottom left} panels indicate the current bound on the gluino mass.}
\label{fig2}
\end{figure}

\begin{figure}[ht!]
\centering
\subfigure{\includegraphics[scale=1.1]{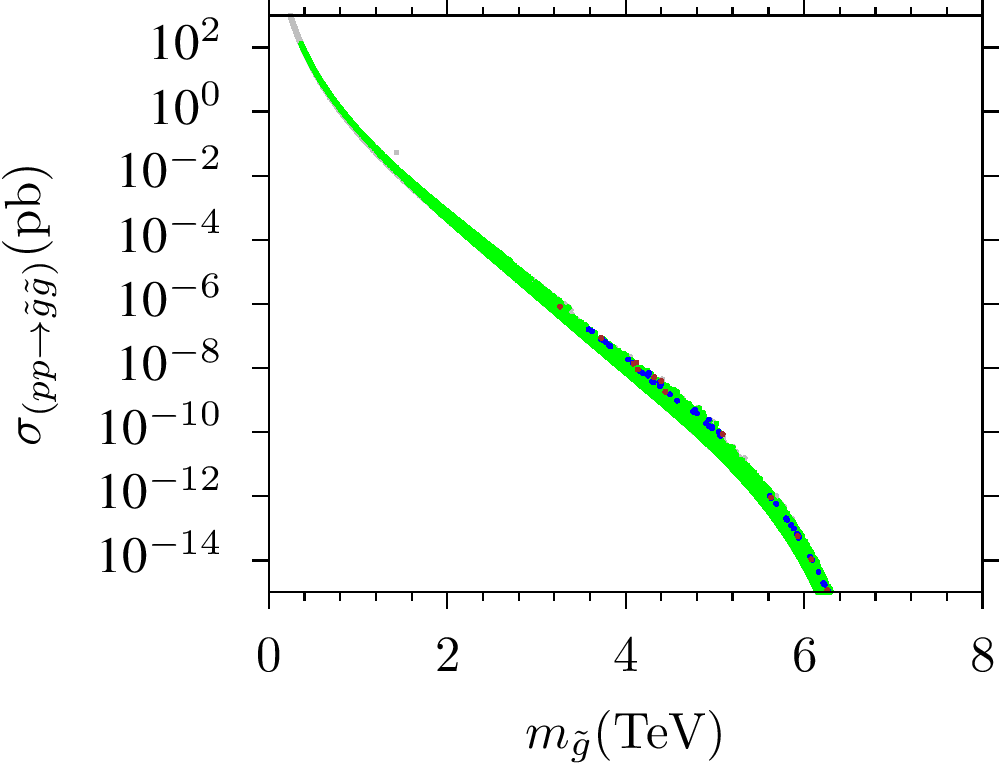}}
\subfigure{\includegraphics[scale=1.1]{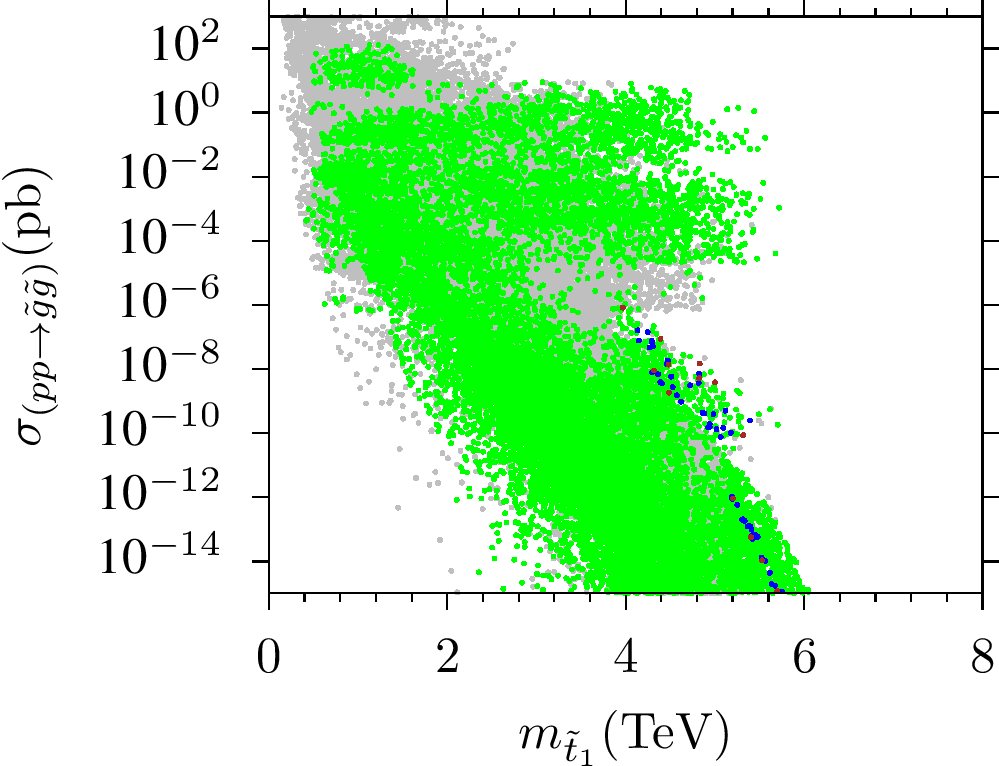}}
\subfigure{\includegraphics[scale=1.1]{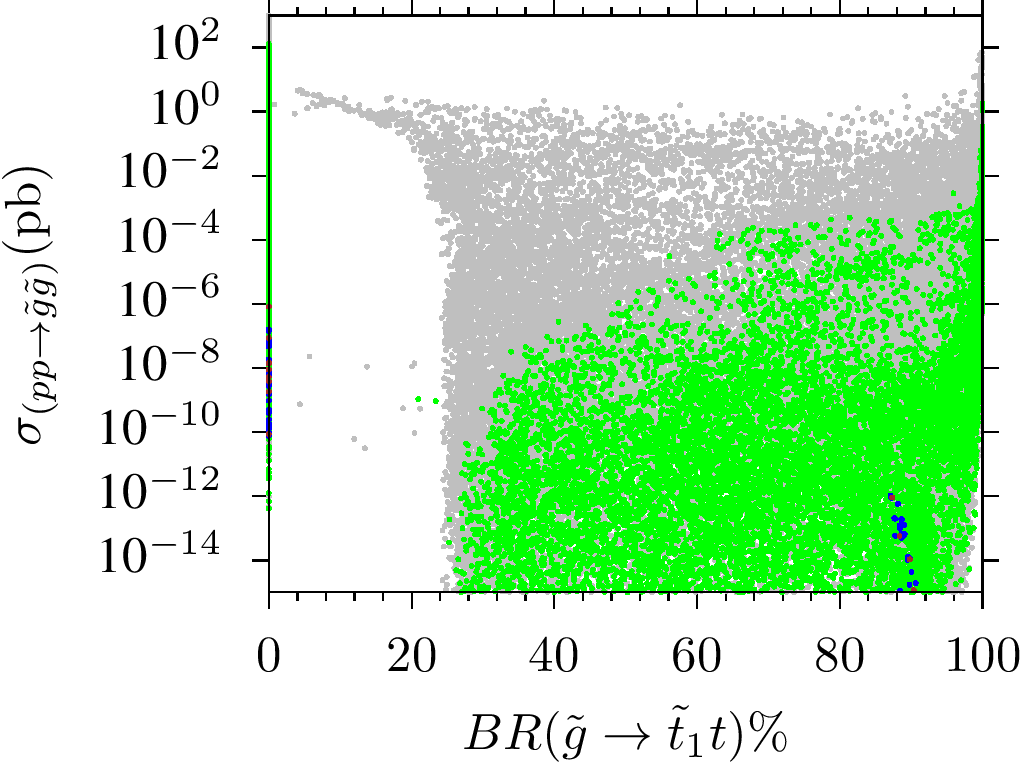}}
\subfigure{\includegraphics[scale=1.1]{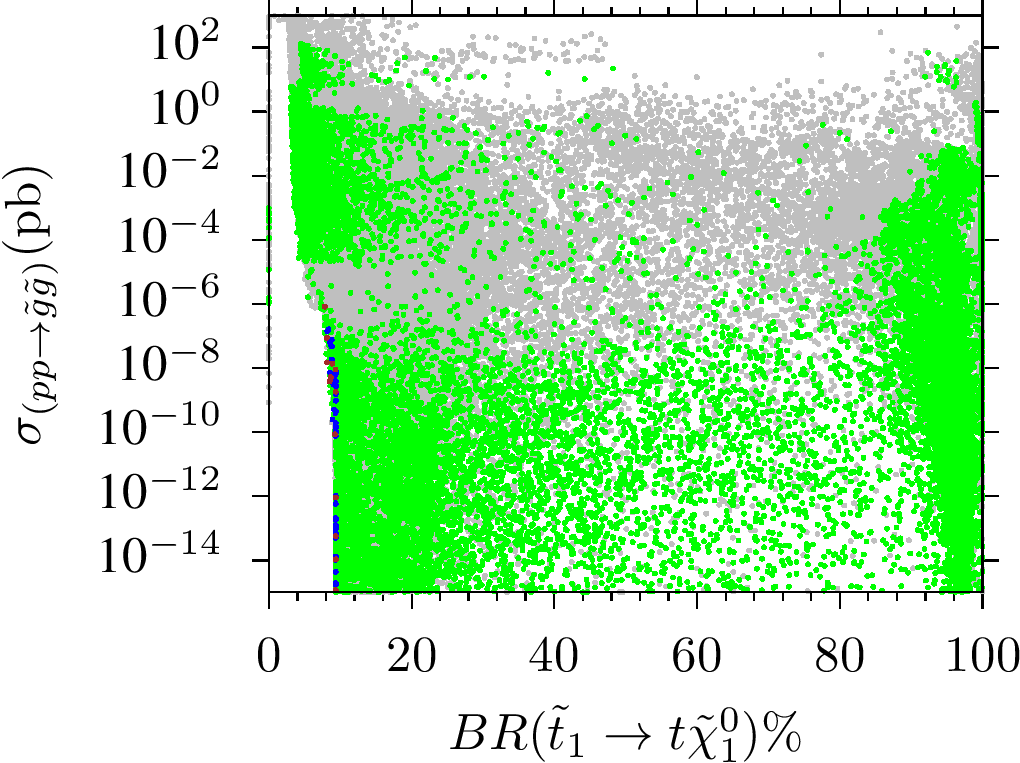}}
\caption{Plots for {the} gluino pair production {cross-section versus}  the gluino {mass} (top left), stop {mass} (top right), ${\rm BR}(\tilde{g}\rightarrow \tilde{t}_{1}t)$ and ${\rm BR}(\tilde{t}_{1}\rightarrow t\tilde{\chi}_{1}^{0})$. The color coding is the same as {in} Figure \ref{fig1}.}
\label{fig3}
\end{figure}

We first discuss the results for the possible signal processes given in Eqs.(\ref{eq:case1}, \ref{eq:case2} and \ref{eq:case3}) in the CMSSM framework, and compare with the current LHC results on the gluino mass scale {to see} if there is any difference. The CMSSM boundary conditions imposed at $M_{{\rm GUT}}$ {yield a} universal SSB mass term $M_{1/2}$ for all MSSM gauginos; hence, $M_{1/2}$ determines both the LSP neutralino and gluino masses. In this context, one expects a linear correlation between the low scale masses of the LSP neutralino and gluino. Figure \ref{fig1} displays our results with plots in the $m_{0}-M_{1/2}$ and $m_{\tilde{g}}-m_{\tilde{t}_{1}}$ planes. All points are compatible with REWSB and neutralino LSP {conditions}. Green points are consistent with the current mass bounds and constraints from rare B-meson decays except {for} gluino, on whose mass the LEP II bound is applied. {Blue} points form a subset of green, and they indicate the solutions allowed by the WMAP bound on the relic abundance of neutralino LSP within $5\sigma$; brown points are a subset of blue and they satisfy the Planck bounds on the relic abundance within $5\sigma$. The horizontal line in the right panel indicates the current mass bound on the gluino mass, and the diagonal line {indicates the mass degeneracy between the stop and gluino}. Even though we allow {relatively low values} for the SSB mass terms for scalars and gauginos, {a} strong bound on these parameters arises from DM {considerations}. The WMAP and Planck bound on the relic abundance of the LSP neutralino (blue and brown points) exclude most of {the} solutions, especially those in regions with $m_{0}\lesssim 5$ TeV and $M_{1/2}\lesssim 1.2$ TeV.

{The low} scale impact of the bounds on these GUT scale parameters can {also be} expressed in terms of gluino and stop masses as shown in the right panel of Figure \ref{fig1}. The  $m_{\tilde{g}}-m_{\tilde{t}_{1}}$ plane shows that {a} gluino heavier than stop {can be realized} in {much} of the parameter space. The DM {considerations} constrain the stop and gluino masses as $m_{\tilde{t}_{1}},m_{\tilde{g}}\gtrsim 4$ TeV, which is beyond the reach of the current LHC experiments at 14 TeV \cite{Baer:2016wkz}, while the near future experiments {provide} some hope to probe {gluino and stop masses in this mass range}.

We show the low scale mass spectrum for gluino, stop, LSP neutralino and the lightest chargino in Figure \ref{fig2} with plots in the $m_{\tilde{g}}-m_{\tilde{\chi}_{1}^{0}}$, ${m_{\tilde{t}_{1}}}-m_{\tilde{\chi}_{1}^{0}}$, $m_{\tilde{g}}-m_{\tilde{\chi}_{1}^{\pm}}$ and ${m_{\tilde{t}_{1}}}-m_{\tilde{\chi}_{1}^{\pm}}$ planes. The color coding is the same as {in} Figure \ref{fig1}. The diagonal lines in the panels {mark the region of equal masses for the particles displayed}. The horizontal lines in the {top left} and  {bottom left} panels indicate the current bound on the gluino mass. As expected and mentioned above, the $m_{\tilde{g}}-m_{\tilde{\chi}_{1}^{0}}$ {plot} reveals a linear correlation between the gluino and the LSP neutralino masses due to the universal gaugino mass term $M_{1/2}$ at $M_{{\rm GUT}}$. As a result, even though we do not apply the current gluino mass bound, CMSSM cannot yield gluino as a next to LSP (NLSP) {particle} in the mass spectrum. On the other hand, {a} gluino as heavy as about 12 TeV {can be realized}. Similarly, the stop can be as heavy as the gluino, while it is also possible to realize the stop as NLSP with $m_{\tilde{t}_{1}}\sim m_{\tilde{\chi}_{1}^{0}} \sim 500$ GeV, as is seen from the $m_{\tilde{t}}-m_{\tilde{\chi}_{1}^{0}}$ plot. As well as neutralino, the lightest chargino can also take part in possible signal processes, {while} the gluino is always much heavier than the chargino as shown in the $m_{\tilde{g}}-m_{\tilde{\chi}_{1}^{\pm}}$ plane. Similarly, the stop is heavier than the chargino in most of the parameter space, while it is also possible to realize $m_{\tilde{t}_{1}}\lesssim m_{\tilde{\chi}_{1}^{\pm}}$ in a small portion of the parameter space (below the diagonal line in the $m_{\tilde{t}_{1}}-m_{\tilde{\chi}_{1}^{\pm}}$ plane). Even though the DM constraints bound the stop and gluino {masses} at about 4 TeV from below, the LSP neutralino and lightest chargino masses are bounded {from below} at about 500 GeV, as seen from all plots in Figure \ref{fig2}.

{The probe of solutions with a gluino lighter than the current bound ($\sim 2.1$ TeV) depends on the strength of the signal,} which is triggered with the gluino pair production, whose cross-section is shown in Figure \ref{fig3} in correlation with the gluino (top left) and stop (top right) masses, ${\rm BR}(\tilde{g}\rightarrow \tilde{t}_{1}t)$ and ${\rm BR}(\tilde{t}_{1}\rightarrow t\tilde{\chi}_{1}^{0})$. The color coding is the same as {in} Figure \ref{fig1}. The $\sigma(p p \rightarrow \tilde{g}\tilde{g})-m_{\tilde{g}}$ plane reveals a decreasing correlation with the gluino mass as it increases. This is expected, since the heavy masses requires greater energies, and the results represented in the $\sigma(p p \rightarrow \tilde{g}\tilde{g})-m_{\tilde{g}}$ plane are consistent with the results previously obtained \cite{Borschensky:2014cia,Baer:2017yqq}. According to our results, the gluino pair production can be realized as high as about 100 pb {if} its mass is about $\mathcal{O}(100{\rm GeV})$, while it drops to about $10^{-2}$ pb {if} the gluino mass is {of order a TeV}. The {DM} constraints lower the cross-section further to about $10^{-6}$ pb for $m_{\tilde{g}}\gtrsim 3.2$ TeV. The production is not likely ($\sigma(pp\rightarrow \tilde{g}\tilde{g}) \ll 10^{-5}$ pb) for $m_{\tilde{g}}\gtrsim 5$ TeV. Even though the correlation is not as sharp as that with the gluino mass, the heavy stop mass scales also exhibit an inverse correlation with the gluino pair production cross-section, as seen in the $\sigma(pp\rightarrow \tilde{g}\tilde{g})-m_{\tilde{t}_{1}}$ plane. Similarly the gluino pair production is not likely for $m_{\tilde{t}_{1}}\gtrsim 5$ TeV. The bottom panels show the possibility of the gluino and stop decays in terms of the branching fractions of the relevant processes. It is possible to find solutions in which ${\rm BR}(\tilde{g}\rightarrow \tilde{t}_{1}t) \approx 100\%$, while the DM {constraints reduce} it down to about $90\%$. The last panel shows how likely the stop can decay into a LSP neutralino along with a top quark, which provides the strongest exclusion on the stop mass. Even though it is possible to realize it as $100\%$, it can only be as high as about $10\%$ when the DM constraints are applied.

\begin{figure}[ht!]
\centering
\subfigure{\includegraphics[scale=1.1]{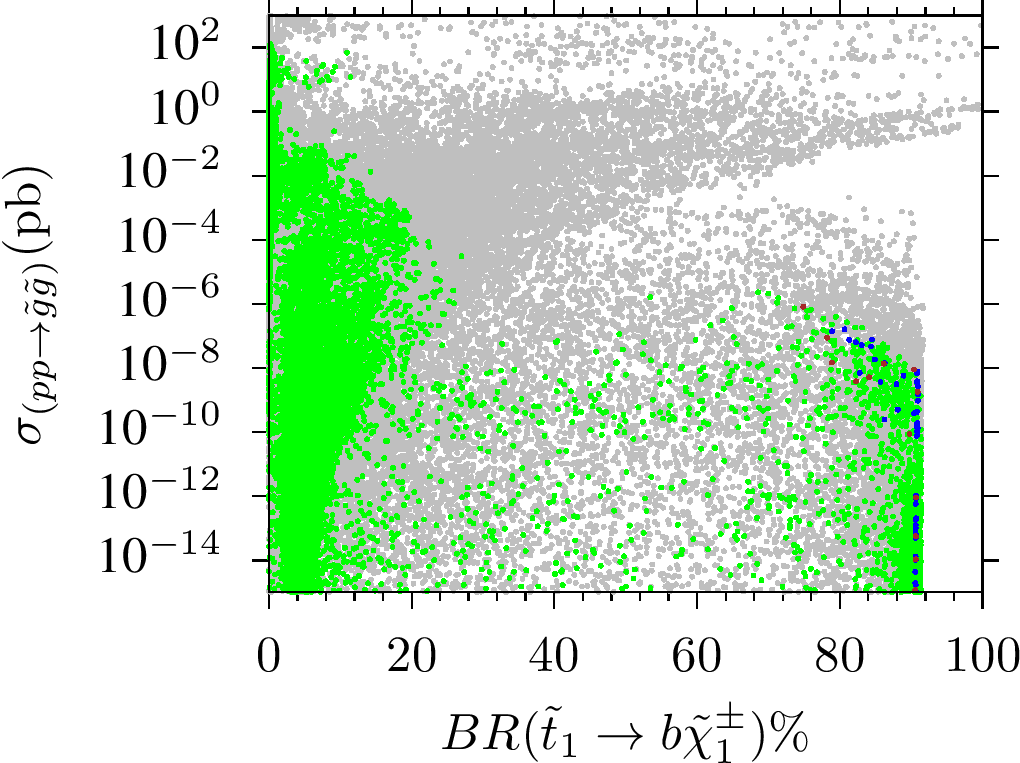}}
\subfigure{\includegraphics[scale=1.1]{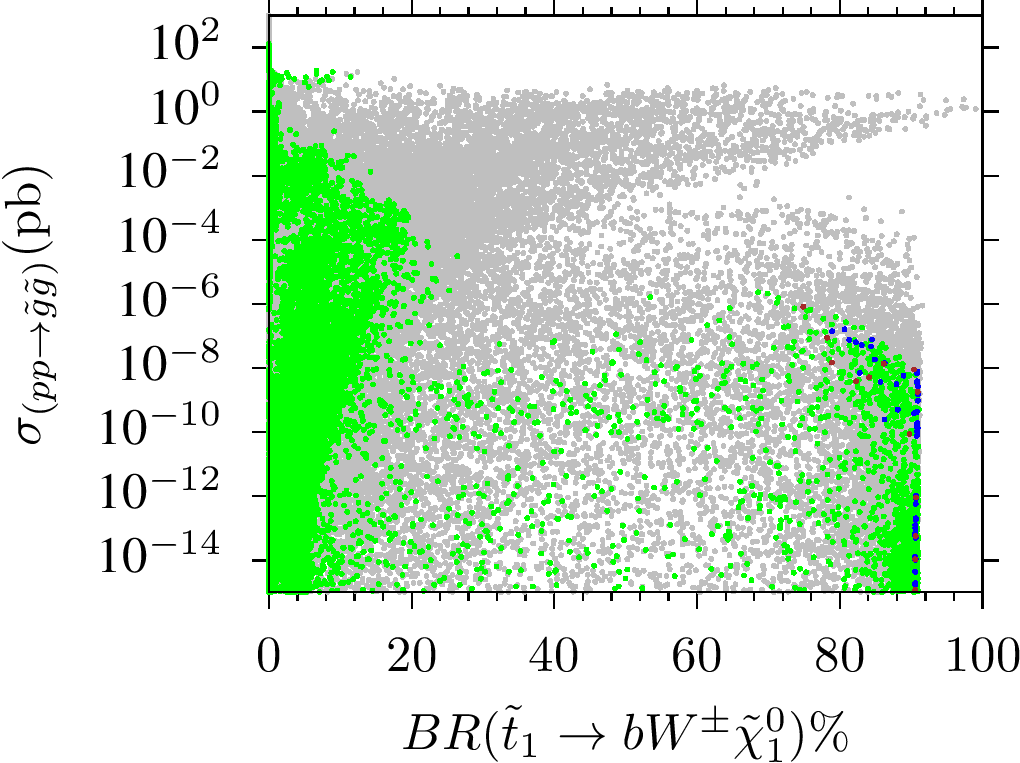}}
\caption{Plots for the gluino pair production {cross-section} and stop decay channels involving bottom and chargino (left), bottom, W-boson and LSP neutralino (right). The color coding is the same as {in} Figure \ref{fig1}.}
\label{fig4}
\end{figure}

Another possibility is {for the stop} to decay into the lightest chargino along with a bottom quark. {In turn, the chargino decays} into the LSP neutralino {and} a $W-$boson. Figure \ref{fig4} represents the results for the cross-section of the gluino pair production and stop decay channels involving bottom and chargino (left), bottom, W-boson and LSP neutralino (right). The color coding is the same as {in} Figure \ref{fig1}. The branching ratio in the right panel is obtained as ${BR}(\tilde{t}_{1}\rightarrow bW^{\pm}\tilde{\chi}_{1}^{0}) \approx {\rm BR}(\tilde{t}_{1}\rightarrow b\tilde{\chi}_{1}^{\pm})\times {\rm BR}(\tilde{\chi}_{1}^{\pm}\rightarrow W^{\pm}\tilde{\chi}_{1}^{0})$. As seen from the panels, {this stop decay mode} can be as high as about $90\%$ {or so consistent} with the DM constraints {for} ${\rm BR}(\tilde{t}_{1}\rightarrow t\tilde{\chi}_{1}^{0})\lesssim 10\%$.

\begin{figure}[ht!]
\centering
\includegraphics[scale=1.5]{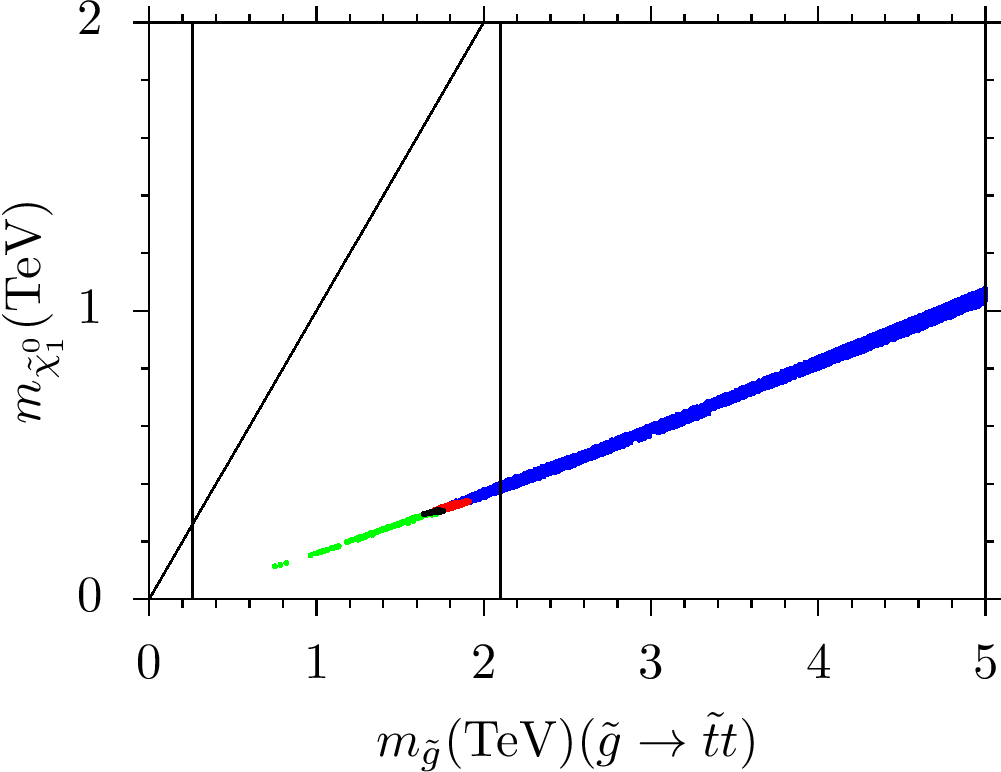}
\caption{{LSP} neutralino and gluino masses in terms of the signal significance for $\tilde{g}\rightarrow \tilde{t}_{1}t$. All points are allowed by the current LHC results listed in Section \ref{sec:scan}. {Blue} points represent the solutions with $SS\gtrsim 0$, the black points show those with $SS\gtrsim 2\sigma$, and red {points} depict those with $SS\gtrsim 1\sigma$. The {remaining solutions in the green region} yield $SS\gtrsim 3\sigma$.}
\label{fig5}
\end{figure}

{Next we consider these} results in terms of the signal strength over the relevant backgrounds which is displayed in Figure \ref{fig5} in the $m_{\tilde{\chi}_{1}^{0}}-m_{\tilde{g}}$ plane. All points are allowed by the current LHC results listed in Section \ref{sec:scan}. The blue points represent the solutions with $SS\gtrsim 0$, while the black points show those with $SS\gtrsim 2\sigma$, and the red {points depict} those with $SS\gtrsim 1\sigma$. The rest of the solutions {lie} in the green region and they yield $SS\gtrsim 3\sigma$. Assuming negative results {for} gluino searches, one can exclude the solutions represented in red, black and green. The blue points yield $SS\gtrsim 0$, which means the $\tilde{g}\rightarrow \tilde{t}_{1}t$ analyses cannot be applied to these points. According to the results shown in Figure \ref{fig5}, one can exclude the gluino mass scales below about 2 TeV {with} $68\%$ CL, while those below about 1.9 TeV can be excluded up to about $95\%$ {CL}. Considering {an error of} about $10\%$ in calculation of the SUSY mass spectrum, these bounds obtained in CMSSM through the gluino pair production, {with the} gluinos decaying into a stop and top quark, are quite similar to those obtained in the low scale analyses.

\begin{figure}[ht!]
\centering
\subfigure{\includegraphics[scale=1.1]{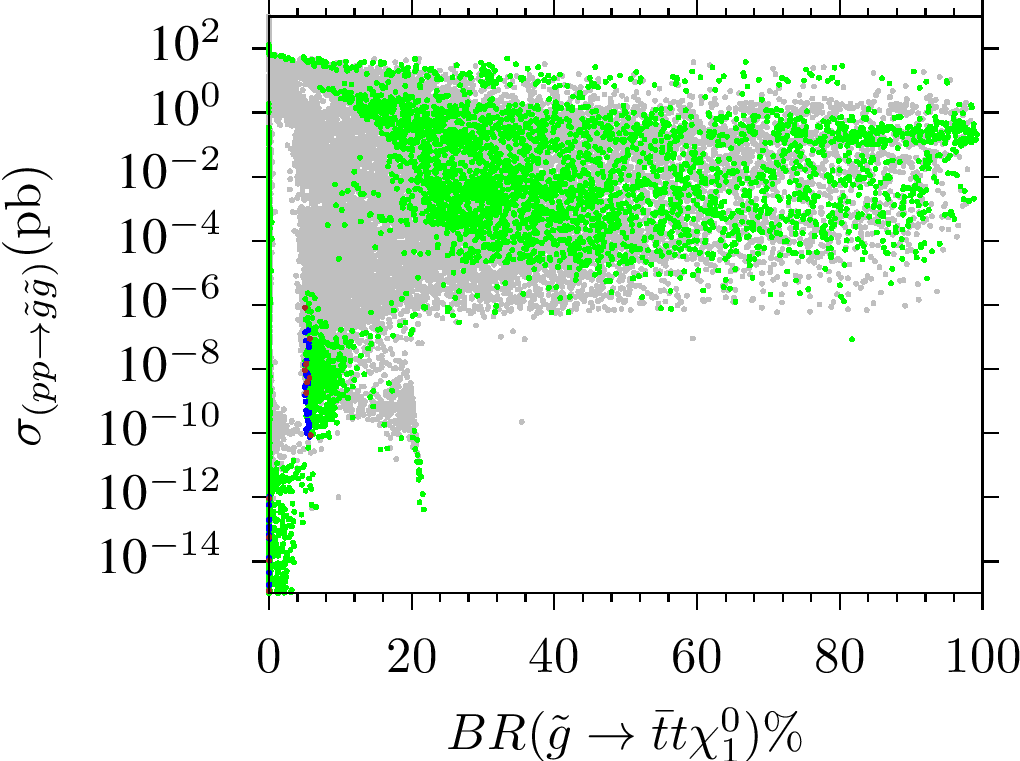}}
\subfigure{\includegraphics[scale=1.1]{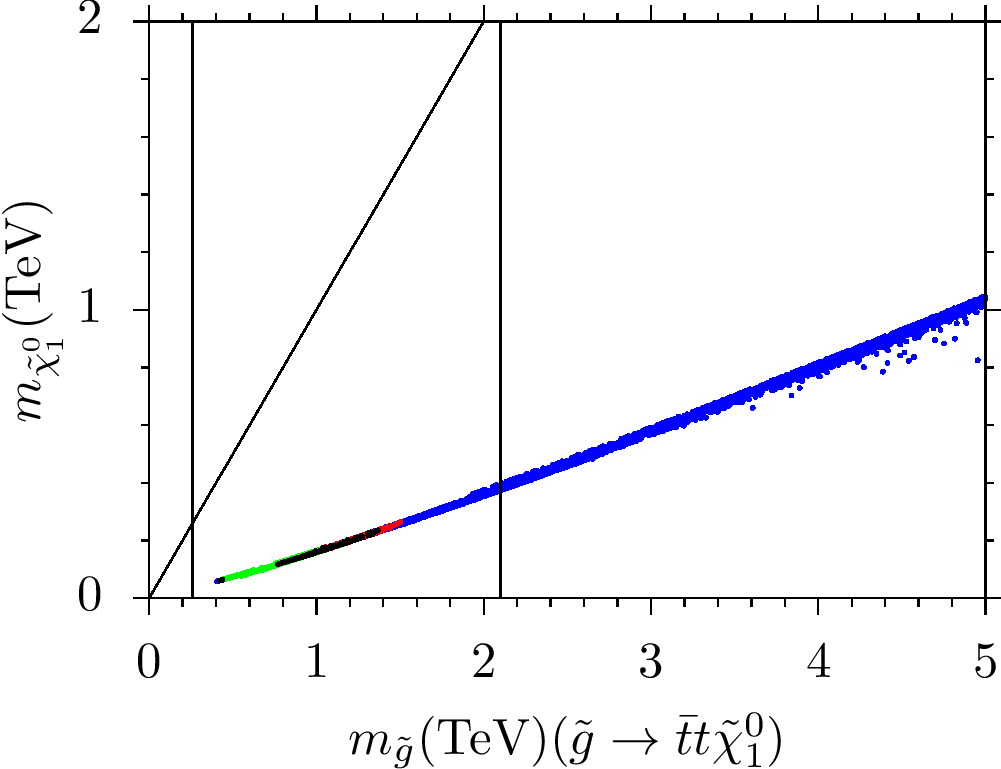}}
\caption{{Plots} in the $\sigma(pp\rightarrow \tilde{g}\tilde{g})-{\rm BR}(\tilde{g}\rightarrow \bar{t}t\tilde{\chi}_{1}^{0})$ and ${m_{\tilde{\chi}_{1}^{0}}-m_{\tilde{g}}}$ planes. The  color coding in the left panel is the same as {in} Figure \ref{fig1}, while the right panel is represented in terms of $SS$, whose color coding is the same as {in} Figure \ref{fig5}.}
\label{fig6}
\end{figure}

Another possibility is {for the gluinos} to decay into a LSP neutralino along with a pair of top quarks, when the $\tilde{g}\rightarrow \tilde{t}_{1}t$ process is not available. Figure \ref{fig6} displays the results for this {case} with the plots in the $\sigma(pp\rightarrow \tilde{g}\tilde{g})-{\rm BR}(\tilde{g}\rightarrow \bar{t}t\tilde{\chi}_{1}^{0})$ and $m_{\tilde{\chi}_{1}^{0}}-m_{\tilde{g}}$ planes. The  color coding in the left panel is the same as {in} Figure \ref{fig1}, while the right panel is represented in terms of $SS$, whose color coding is the same as {in} Figure \ref{fig5}. The exclusion {region for} the gluino mass scale can be slightly lowered through this channel as the regions with $m_{\tilde{g}}\lesssim 1.6$ TeV are excluded up to about $68\%$ {CL}, while those with $m_{\tilde{g}}\lesssim 1.4$ TeV are excluded up to about $95\%$ {CL}.

\section{Gluino Exclusion in NUGM}
\label{sec:NUGM}

\begin{figure}[ht!]
\centering
\subfigure{\includegraphics[scale=1.1]{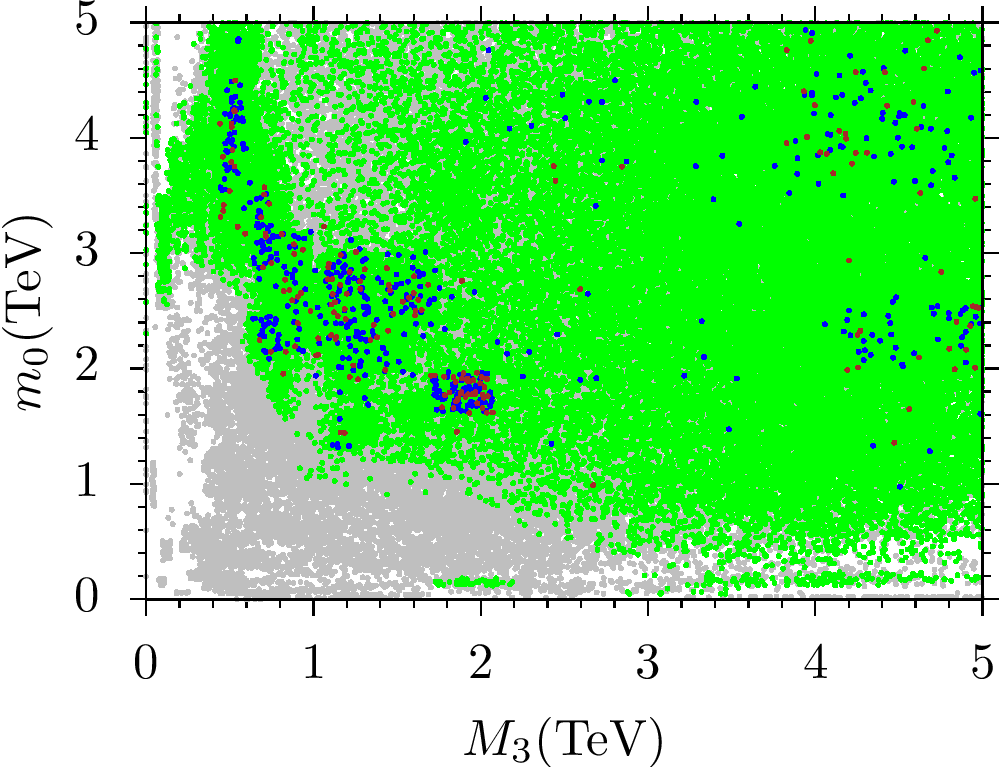}}
\subfigure{\includegraphics[scale=1.1]{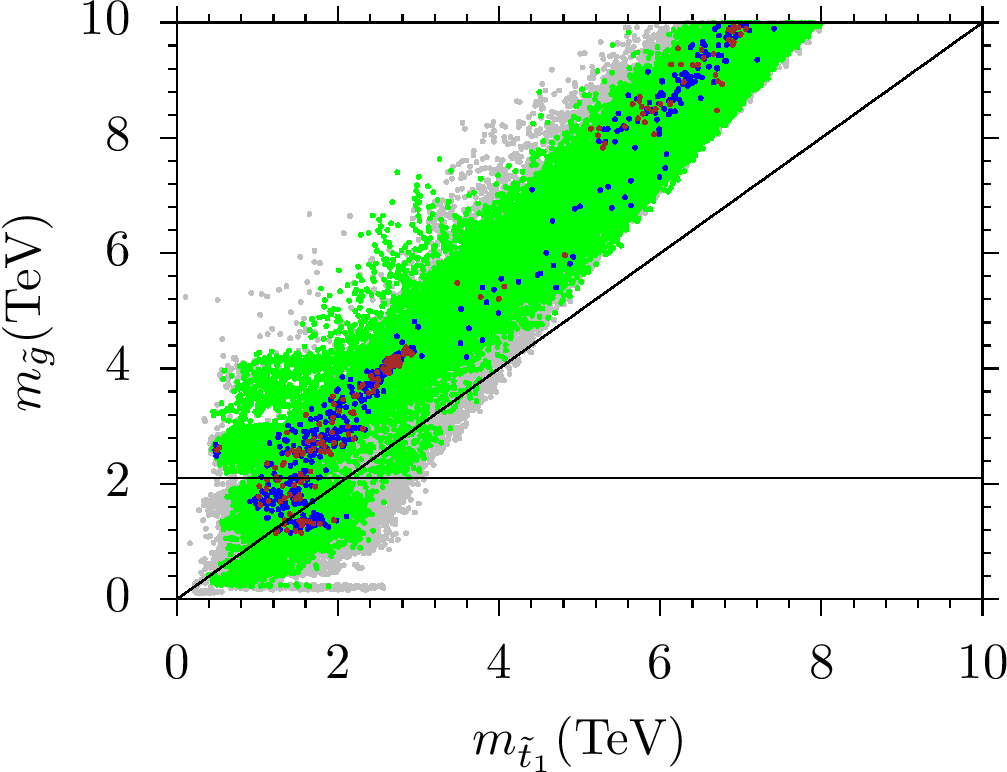}}
\caption{Plots in the $m_{0}-M_{3}$ and $m_{\tilde{g}}-m_{\tilde{t}_{1}}$ planes. The color coding is the same as {in} Figure \ref{fig1}. The diagonal line in the right panel indicates the mass degeneracy between the gluino and stop.} 
\label{fig:NUGM1}
\end{figure}

In this section we relax the universality in the gaugino sector by setting all gaugino masses independent {of} each other such that the linear correlation between the LSP neutralino and gluino masses does not hold. Non-universality in SUSY GUTs can be realized in several cases without conflicting with the underlying GUT symmetry and gauge coupling unification {using} non-singlet $F-$terms with non-zero VEVs \cite{Martin:2009ad}, or $F-$terms which are a linear combination of two or more distinct fields from different representations \cite{Martin:2013aha}. Another way is to assume two distinct sources of SUSY breaking \cite{Anandakrishnan:2013cwa}. 

We first display our results for the scalar and gaugino masses in terms of the GUT parameters and low scale masses of the stop and gluino in Figure \ref{fig:NUGM1} in the $m_{0}-M_{3}$ and $m_{\tilde{g}}-m_{\tilde{t}_{1}}$ planes. The color coding is the same as {in} Figure \ref{fig1}. The diagonal line in the right panel indicates the mass degeneracy between the gluino and stop. As seen from the $m_{0}-M_{3}$ plane, the DM constraints do not exclude any region in the fundamental parameter space, even though some regions in gray are excluded by the current LHC results. However, the correlation between the gluino and stop masses still holds {as} seen from the $m_{\tilde{g}}-m_{\tilde{t}_{1}}$ {plane}, since these SUSY particles raise mutually their masses in the RGEs. The $m_{\tilde{g}}-m_{\tilde{t}_{1}}$ plane shows that {the} gluino happens {to be} heavier than {the} stop in most of the parameter space, while a {relatively} small portion can yield {a} stop heavier than {the gluino}. Moreover, the DM constraints allow the gluino and stop to be as low as about 1 TeV, in contrast to {the CMSSM case}.

\begin{figure}[ht!]
\centering
\subfigure{\includegraphics[scale=1.1]{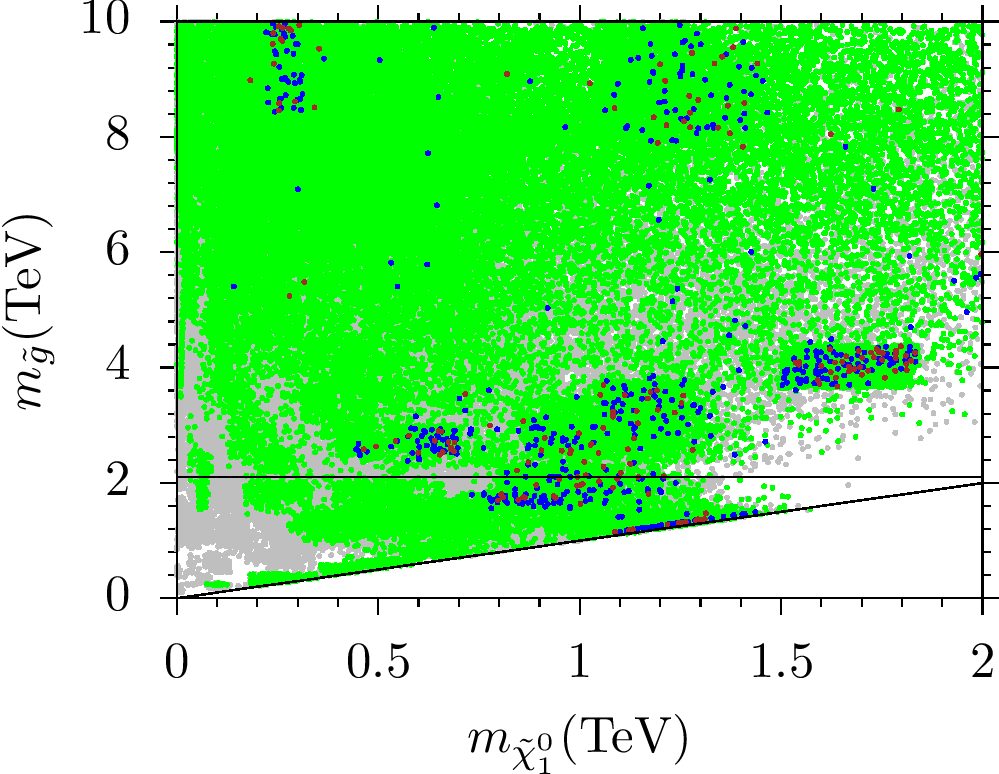}}
\subfigure{\includegraphics[scale=1.1]{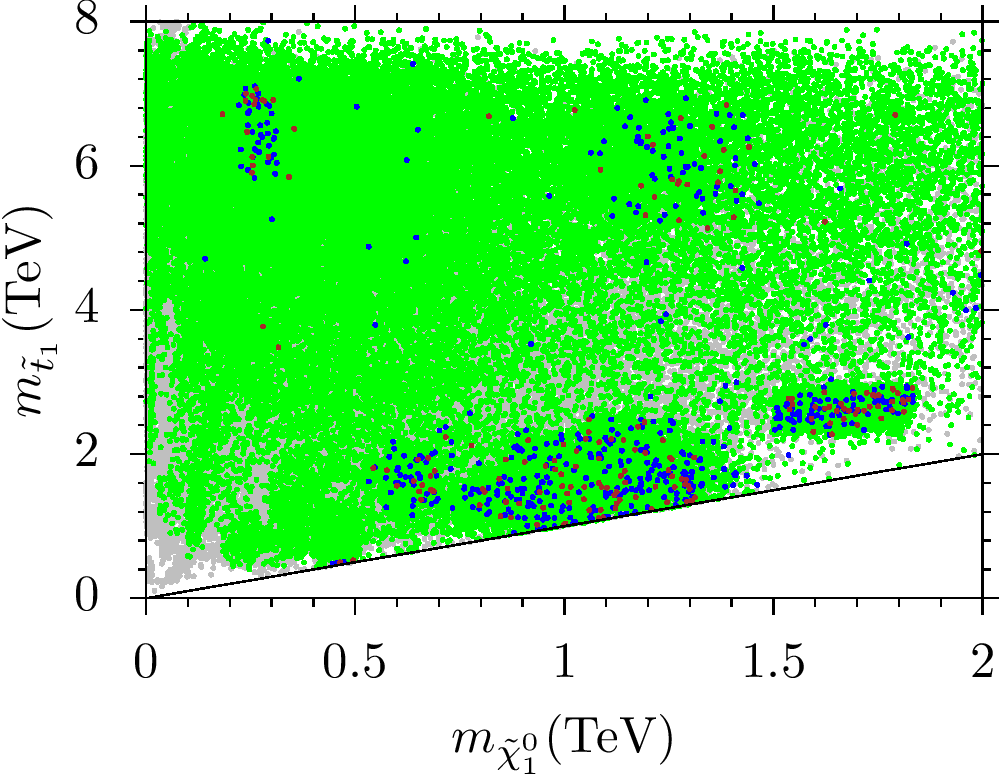}}
\subfigure{\includegraphics[scale=1.1]{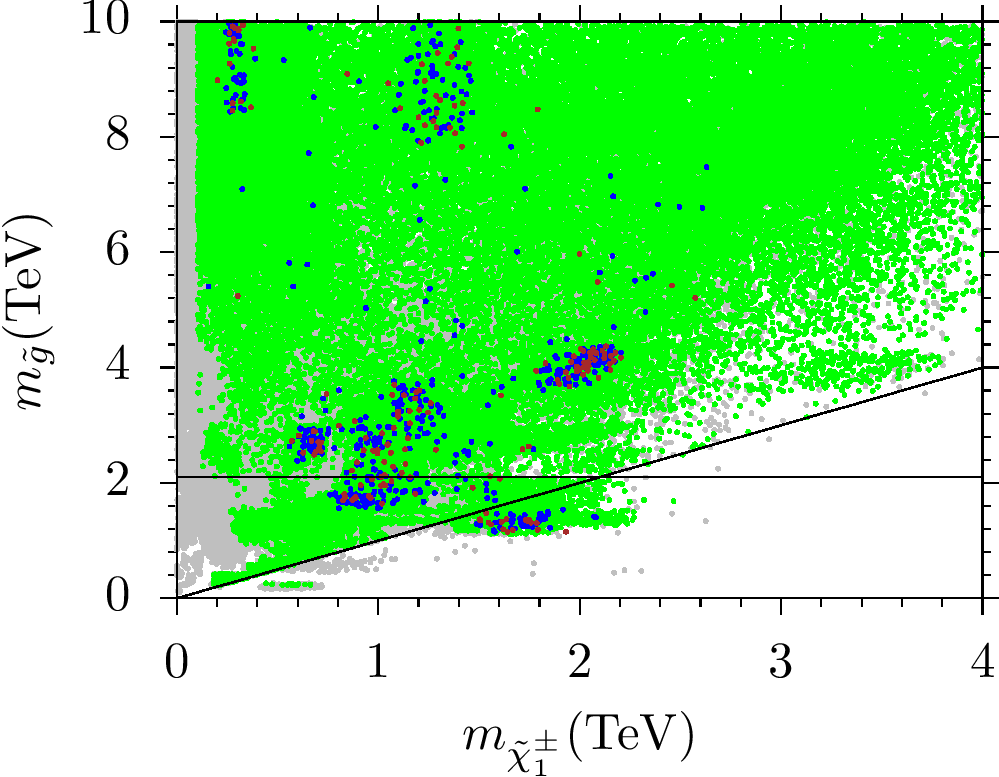}}
\subfigure{\includegraphics[scale=1.1]{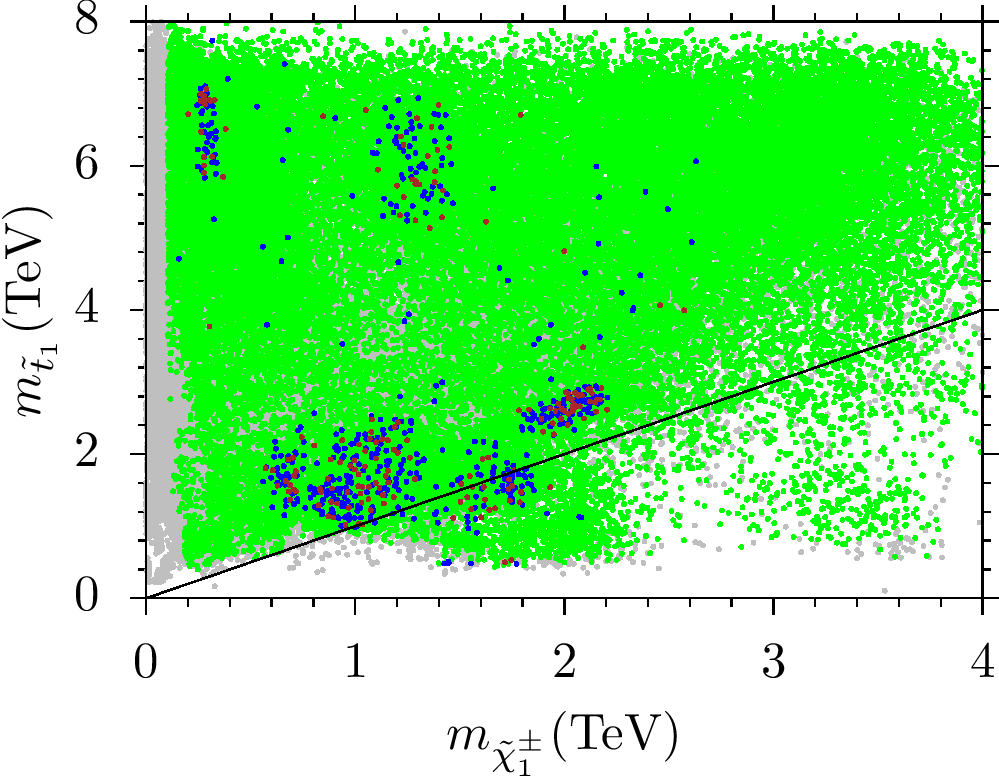}}
\caption{Plots for the mass spectrum in the $m_{\tilde{g}}-m_{\tilde{\chi}_{1}^{0}}$, $m_{\tilde{t}_{1}}-m_{\tilde{\chi}_{1}^{0}}$, $m_{\tilde{g}}-m_{\tilde{\chi}_{1}^{\pm}}$ and ${m_{\tilde{t}_{1}}-m_{\tilde{\chi}_{1}^{\pm}}}$ planes. The color coding is the same as {in} Figure \ref{fig1}, and the diagonal lines indicate the {degeneracy between masses shown.}}
\label{fig:NUGM2}
\end{figure}

We discuss the low scale mass spectrum in more details with plots in the $m_{\tilde{g}}-m_{\tilde{\chi}_{1}^{0}}$, $m_{\tilde{t}_{1}}-m_{\tilde{\chi}_{1}^{0}}$, $m_{\tilde{g}}-m_{\tilde{\chi}_{1}^{\pm}}$ and ${m_{\tilde{t}_{1}}-m_{\tilde{\chi}_{1}^{\pm}}}$ planes of Figure \ref{fig:NUGM2}. The color coding is the same as {in} Figure \ref{fig1}, and the diagonal lines indicate the mass degeneracy between the particles {involved}. The $m_{\tilde{g}}-m_{\tilde{\chi}_{1}^{0}}$ plane reveals that the gluino {is} nearly degenerate with {the} LSP neutralino for mass scales $1.1 \gtrsim m_{\tilde{g}} \gtrsim 1.5$ TeV. Such solutions favor the gluino-neutralino coannihilation {scenarios} which are able to {bring} the {thermal} relic abundance of the LSP neutralino to the ranges allowed by the WMAP and Planck measurements. Besides, since $m_{\tilde{g}}\approx m_{\tilde{\chi}_{1}^{0}}$, these solutions {forbid processes} $\tilde{g}\rightarrow \tilde{t}_{1}t$ and $\tilde{g}\rightarrow \bar{t}t\tilde{\chi}_{1}^{0}$. Indeed, the exclusion {limit} on the gluino mass is not severe ($m_{\tilde{g}} \gtrsim 800$ GeV) {if} it happens to be {the} NLSP. Similarly, we can also identify the stop-neutralino coannihilation solutions {with} $0.9 \gtrsim m_{\tilde{t}_{1}}\approx m_{\tilde{\chi}_{1}^{0}} \gtrsim 1.5$ TeV, around the diagonal line in the $m_{\tilde{t}_{1}}-m_{\tilde{\chi}_{1}^{0}}$ {plot}. {For} these solutions, the constraint on the relic abundance of the LSP neutralino is satisfied through stop-neutralino coannihilation scenario. The bottom panels of Figure \ref{fig:NUGM2} show the chargino {mass compared to} gluino (left) and stop (right). Although the chargino can be as heavy as about 3 TeV, it is always lighter than the gluino and stop, unless the gluino and/or stop are NLSP.

\begin{figure}[ht!]
\subfigure{\includegraphics[scale=1.1]{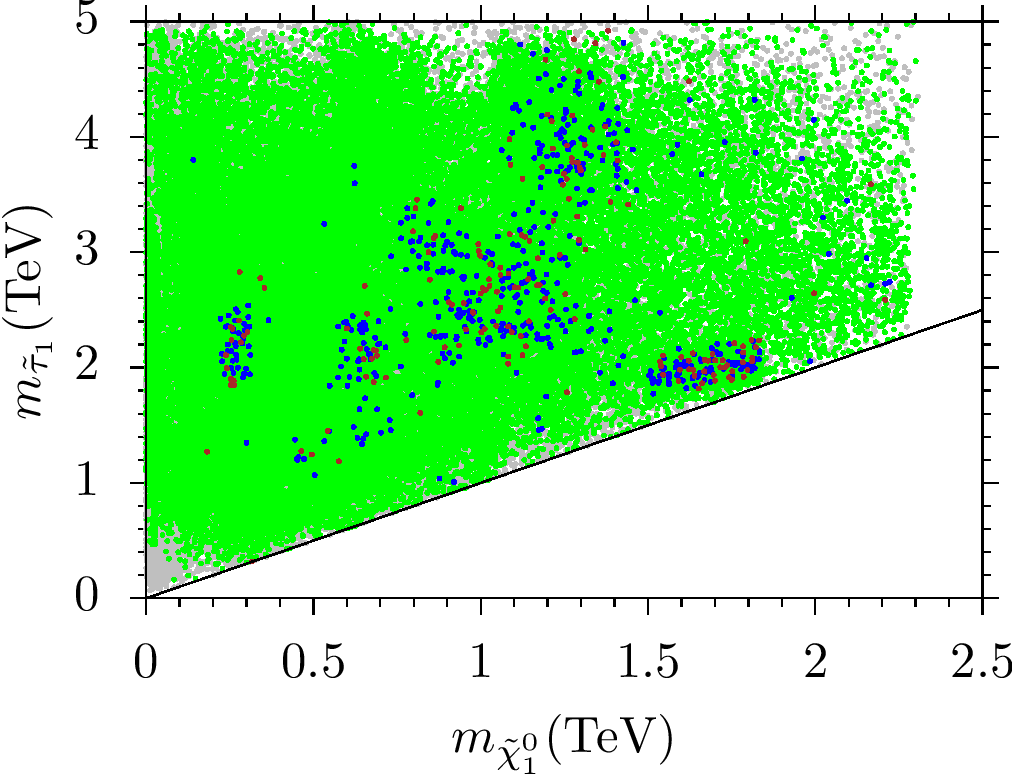}}
\subfigure{\includegraphics[scale=1.1]{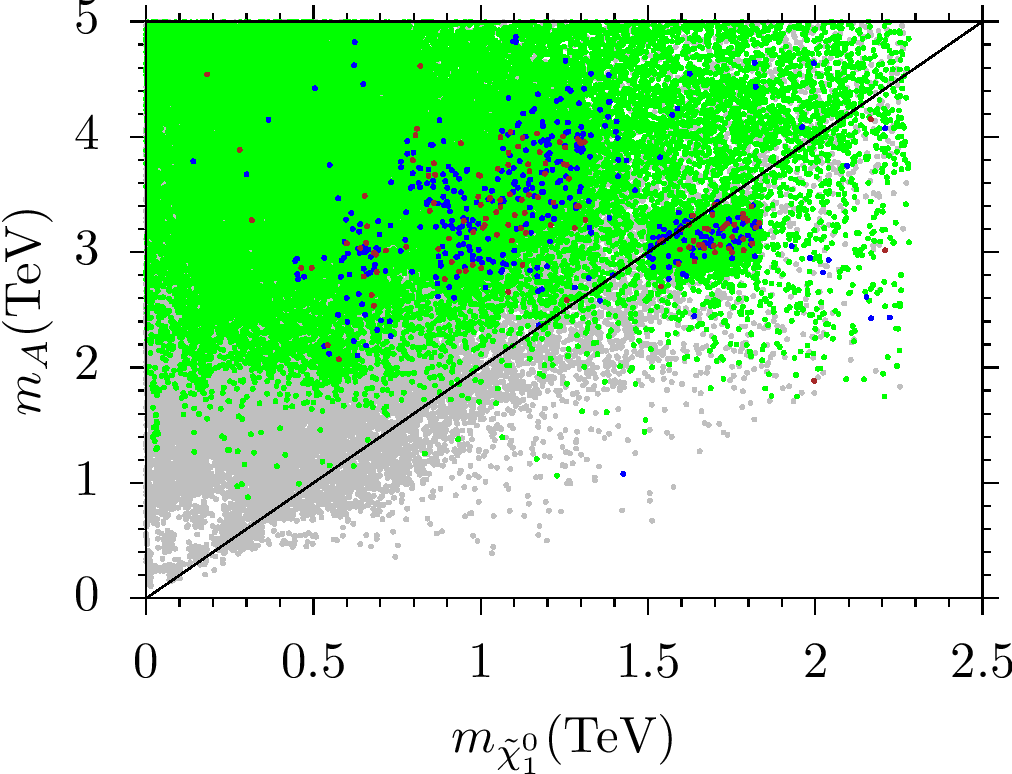}}
\caption{{Mass} spectrum in the $m_{\tilde{\tau}_{1}}-m_{\tilde{\chi}_{1}^{0}}$ and $m_{A}-m_{\tilde{\chi}_{1}^{0}}$ planes. The color coding is the same as {in} Figure \ref{fig1}. The diagonal line in the left panel {displays} the {region} with $m_{\tilde{\tau}_{1}}=m_{\tilde{\chi}_{1}^{0}}$, while it represents $m_{A}=2m_{\tilde{\chi}_{1}^{0}}$ in the right panel.}
\label{fig:NUGM_DM}
\end{figure}

In addition to gluino and stop, we present the mass spectrum in the $m_{\tilde{\tau}_{1}}-m_{\tilde{\chi}_{1}^{0}}$ and $m_{A}-m_{\tilde{\chi}_{1}^{0}}$ planes. The color coding is the same as {in} Figure \ref{fig1}. The diagonal line in the left panel indicates the regions with $m_{\tilde{\tau}_{1}}=m_{\tilde{\chi}_{1}^{0}}$, while it represents $m_{A}=2m_{\tilde{\chi}_{1}^{0}}$ in the right panel. {It} shows that the stau can be as heavy as about 2.2 TeV, and it is degenerate with the LSP neutralino {to} within $10 \%$ {for} $1.5 \lesssim {m_{\tilde{\tau}_{1}}\sim m_{\tilde{\chi}_{1}^{0}} } \lesssim 1.9$ TeV. The solutions in this mass scale favor the stau-neutralino coannihilation scenario in satisfying the constraint on the relic abundance of the LSP neutralino. This mass scale also favors {the} $A-$ resonance {solution} as shown in the $m_{A}-m_{\tilde{\chi}_{1}^{0}}$ plane.

\begin{figure}[ht!]
\centering
\subfigure{\includegraphics[scale=1.1]{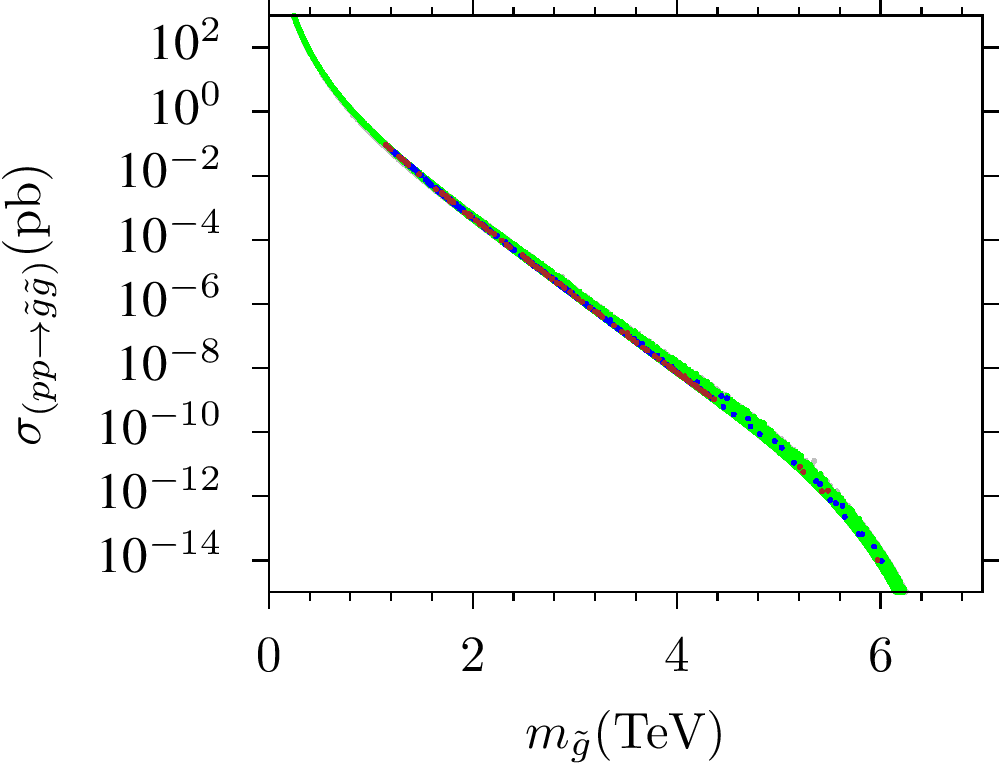}}
\subfigure{\includegraphics[scale=1.1]{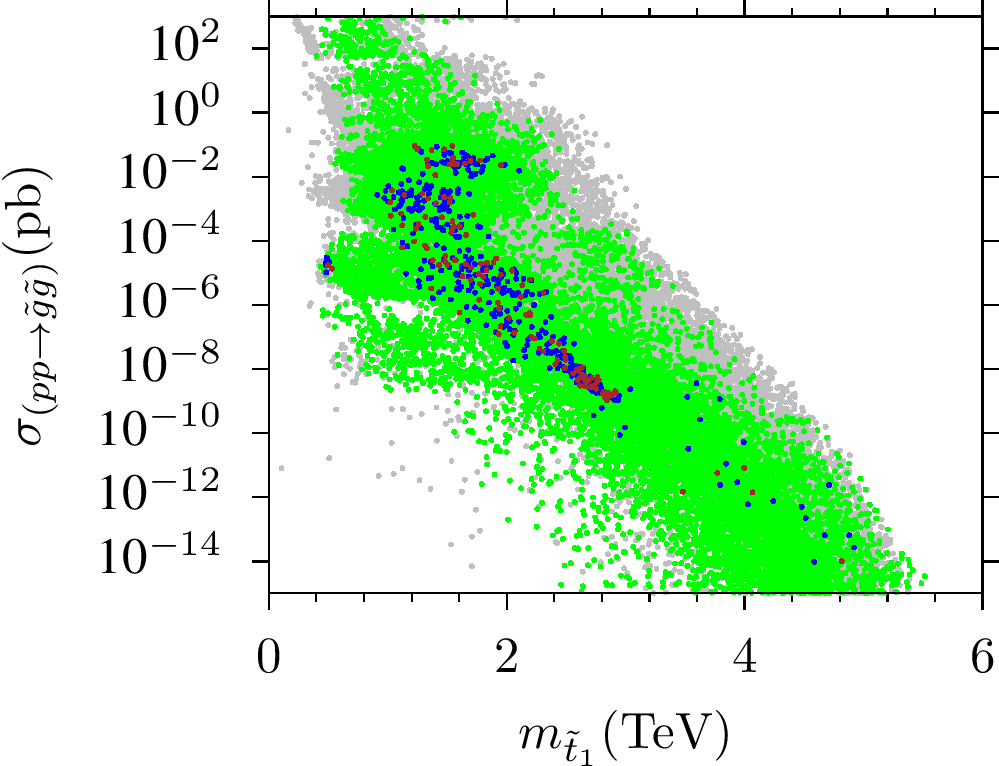}}
\subfigure{\includegraphics[scale=1.1]{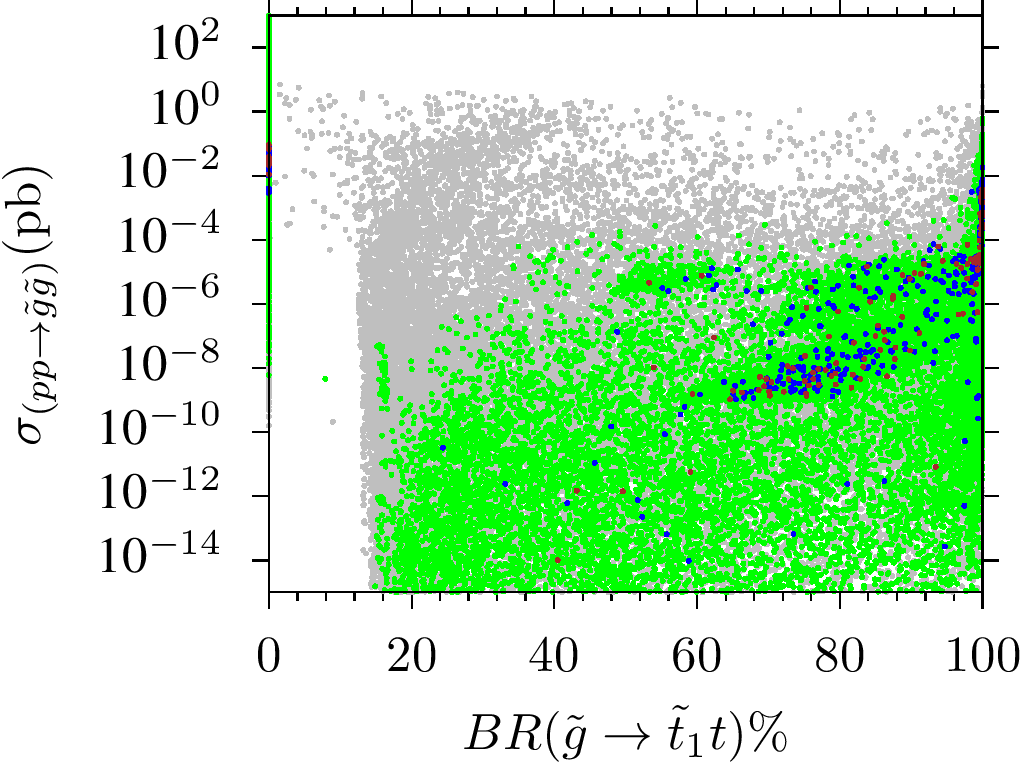}}
\subfigure{\includegraphics[scale=1.1]{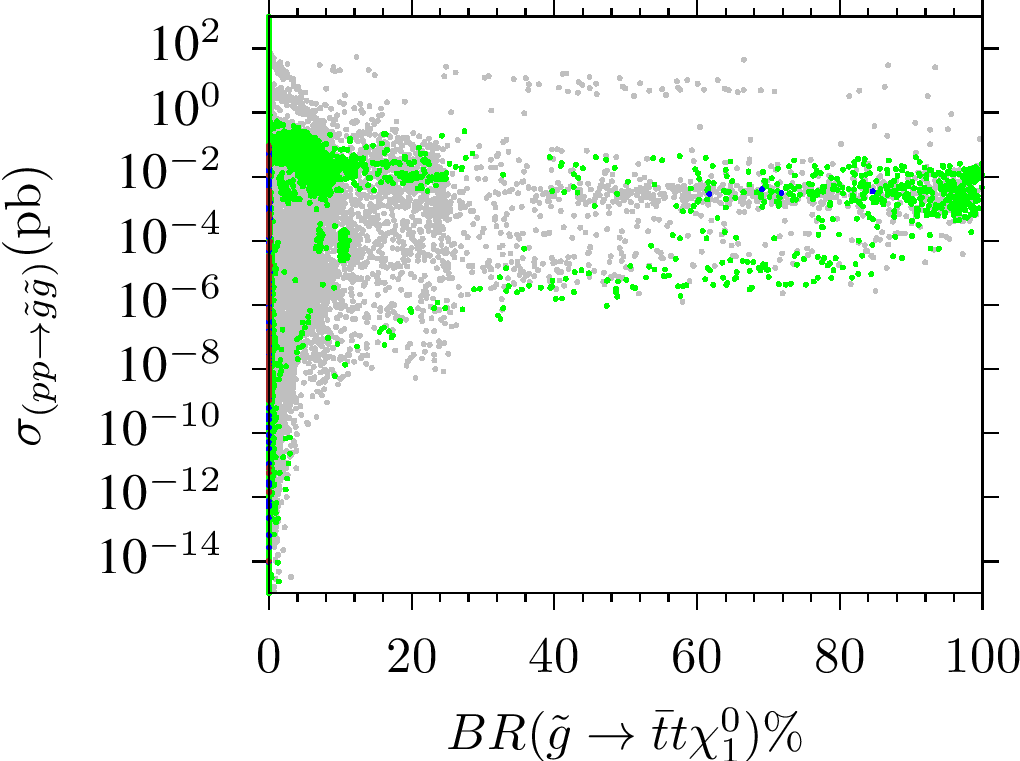}}
\caption{Plots for {gluino} pair production cross-section and its decay modes in the $\sigma(pp\rightarrow \tilde{g}\tilde{g})- m_{\tilde{g}}$, $\sigma(pp\rightarrow \tilde{g}\tilde{g})-m_{\tilde{t}_{1}}$, $\sigma(pp\rightarrow \tilde{g}\tilde{g})-{\rm BR}(\tilde{g}\rightarrow \tilde{t}_{1}t)$ and ${\rm BR}(\tilde{g}\rightarrow \bar{t}t\tilde{\chi}_{1}^{0})$ {planes}. The color coding is the same as {in} Figure \ref{fig1}.}
\label{fig:NUGM3}
\end{figure}

Figure \ref{fig:NUGM3} shows our results for the gluino pair production cross section at 14 TeV and its decay modes in the $\sigma(pp\rightarrow \tilde{g}\tilde{g})- m_{\tilde{g}}$, $\sigma(pp\rightarrow \tilde{g}\tilde{g})-m_{\tilde{t}_{1}}$, $\sigma(pp\rightarrow \tilde{g}\tilde{g})-{\rm BR}(\tilde{g}\rightarrow \tilde{t}_{1}t)$ and ${\rm BR}(\tilde{g}\rightarrow \bar{t}t\tilde{\chi}_{1}^{0})$ {planes}. The color coding is the same as {in} Figure \ref{fig1}. The results are quite similar to those obtained in the CMSSM framework. This is because the particle dynamics remain the same, since we do not extend the particle content and/or the symmetry group. The gluino pair production cross-section can be {of} the order $\mathcal{O}(100)$ pb for light gluino solutions, while the DM constraints lower it to about $10^{-1}$ pb by bounding the gluino mass at about 1 TeV {from below}. The {cross-section} drops below $10^{-5}$ pb for $m_{\tilde{g}}\gtrsim 3$ TeV, beyond which the gluino pair production becomes unlikely. Similar discussion can be followed {for the stop mass} considered as shown in the $\sigma(pp\rightarrow \tilde{g}\tilde{g})-m_{\tilde{t}_{1}}$ {plot}. The $\sigma(pp\rightarrow \tilde{g}\tilde{g})-{\rm BR}(\tilde{g}\rightarrow \tilde{t}_{1}t)$ plane shows that the gluino can decay $100\%$ into a stop and top quark consistent with the LHC constraints and DM measurements. The $\tilde{g}\rightarrow \bar{t}t\tilde{\chi}_{1}^{0}$ process can also be realized at $100\%$, but the DM constraints {mostly} exclude this process. We should note that there are a few solutions satisfying the DM constraints {for} ${\rm BR}(\tilde{g}\rightarrow \bar{t}t\tilde{\chi}_{1}^{0}) \lesssim 85\%$. The number of such solutions can {increase in a more} thorough statistical distribution\footnote{Our distribution is also poor in regard {to} analyzing Signal\ref{eq:case2}, but based on the results from CMSSM, we can expect {an exclusion similar} to that obtained in the case of Signal\ref{eq:case1}.}.

\begin{figure}[ht!]
\centering
\subfigure[14 TeV]{\includegraphics[scale=1.1]{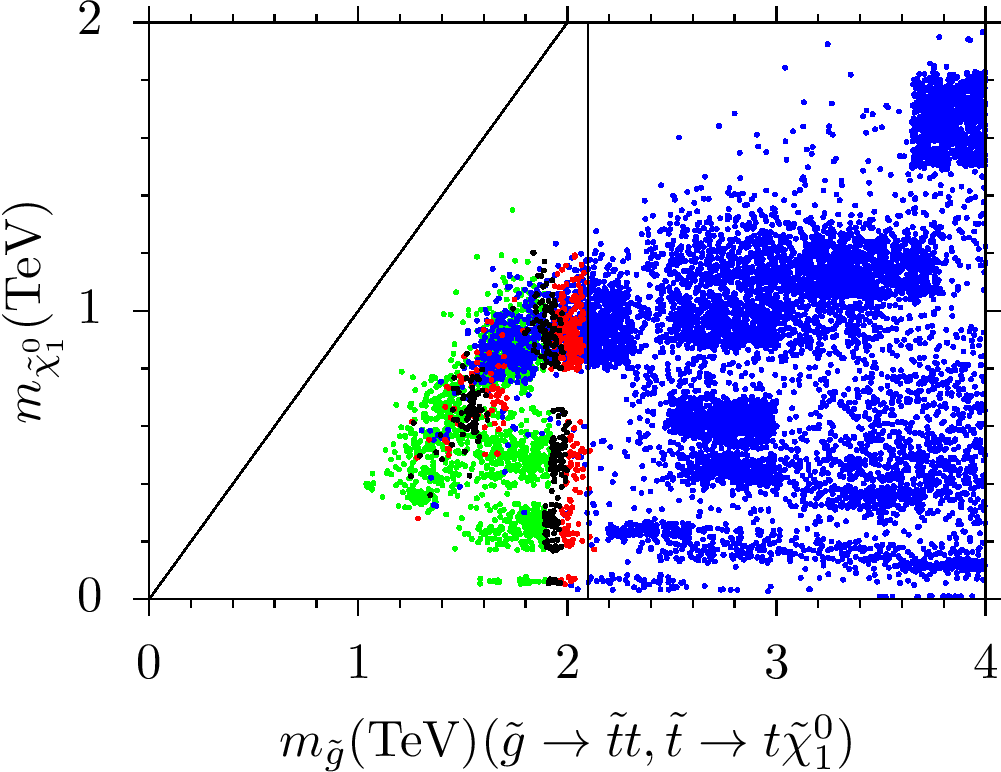}}
\subfigure[27 TeV]{\includegraphics[scale=1.1]{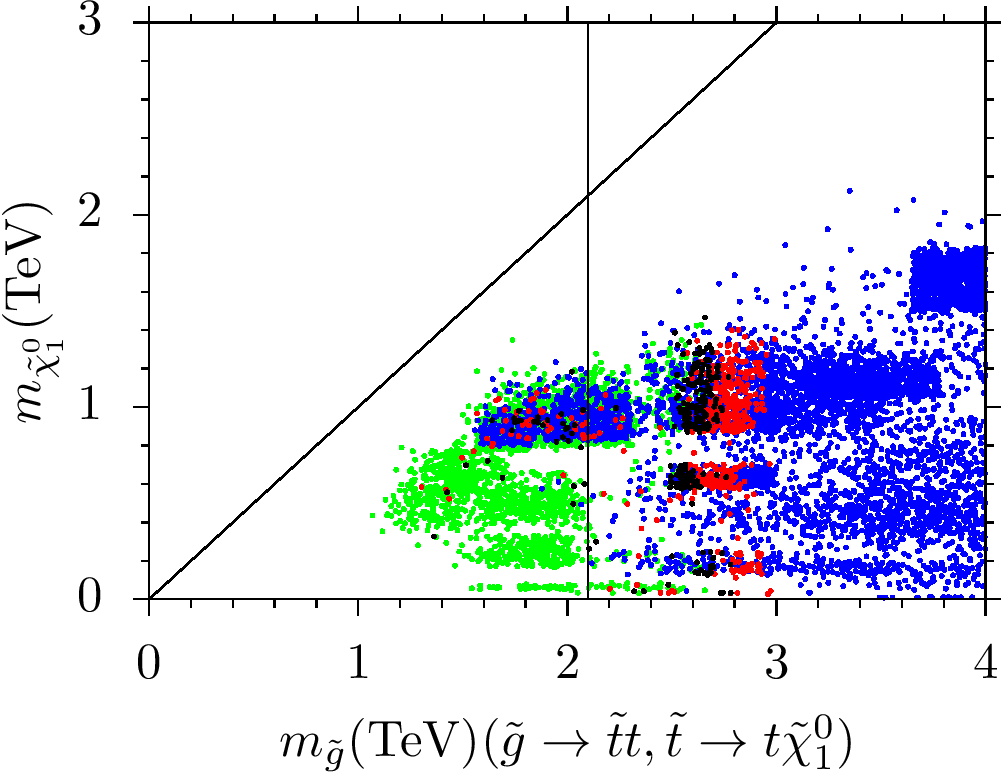}}
\subfigure{\includegraphics[scale=1.1]{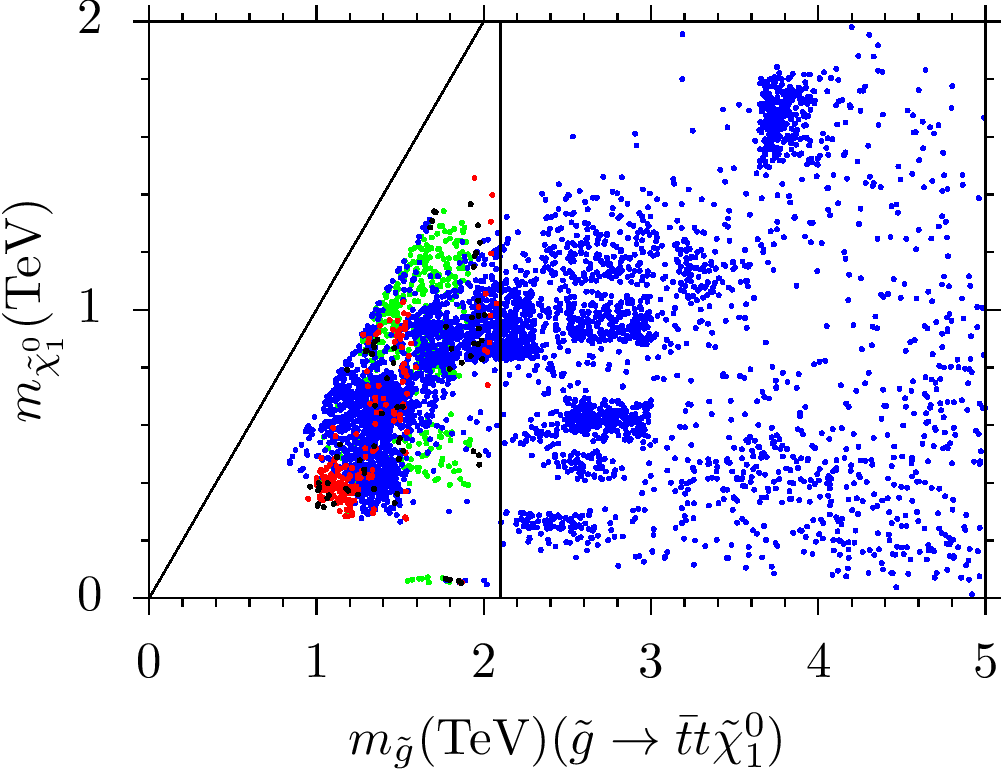}}
\subfigure{\includegraphics[scale=1.1]{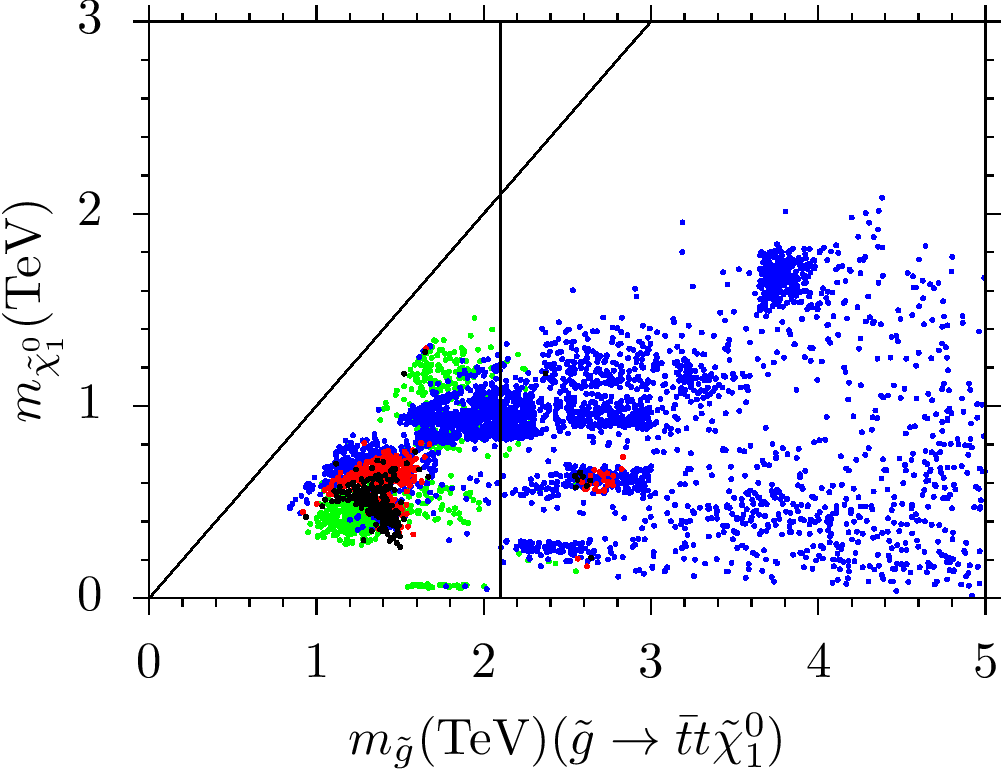}}
\caption{{LSP} neutralino and gluino masses in terms of the signal significance for the $\tilde{g}\rightarrow \tilde{t}_{1}t$ and $\tilde{g}\rightarrow \bar{t}t\tilde{\chi}_{1}^{0}$. The color coding is the same as {in} Figure \ref{fig5}.}
\label{fig:NUM4}
\end{figure}

Assuming negative results in {search} of the gluino and/or stop at 14 and 27 TeV, the excluded regions are shown for the $\tilde{g}\rightarrow \tilde{t}_{1}t$ and $\tilde{g}\rightarrow \bar{t}t\tilde{\chi}_{1}^{0}$ in the LSP neutralino and gluino mass plane of Figure \ref{fig:NUM4} in terms of the signal significance. The color coding is the same as {in} Figure \ref{fig5}. The top left panel reveals {exclusion plots quite similar} to those from the ATLAS and CMS experiments {with a gluino mass scale} lower than about 2 TeV excluded {if} the gluino decays into a stop and a top quark. This {procedure} can probe the gluino {mass} up to about 3 TeV {if} the center of mass energy is raised to 27 GeV, as seen from the top right panel. The analyzes for the case of Signal\ref{eq:case3} {also} yields similar exclusion {plots} as shown in the bottom panels. {Gluino} masses below about 2.1 TeV are excluded for 14 TeV, while 27 TeV can exclude regions with $m_{\tilde{g}}\lesssim 3$ TeV. Note that the NLSP gluino solutions have disappeared from the planes of Figure \ref{fig:NUM4} due to the conditions $\sigma({\rm Signal\ref{eq:case1}}) > 0$ in the top panels, and $\sigma({\rm Signal\ref{eq:case3}}) > 0$ in the bottom panels.

\begin{figure}[ht!]
\subfigure{\includegraphics[scale=1.1]{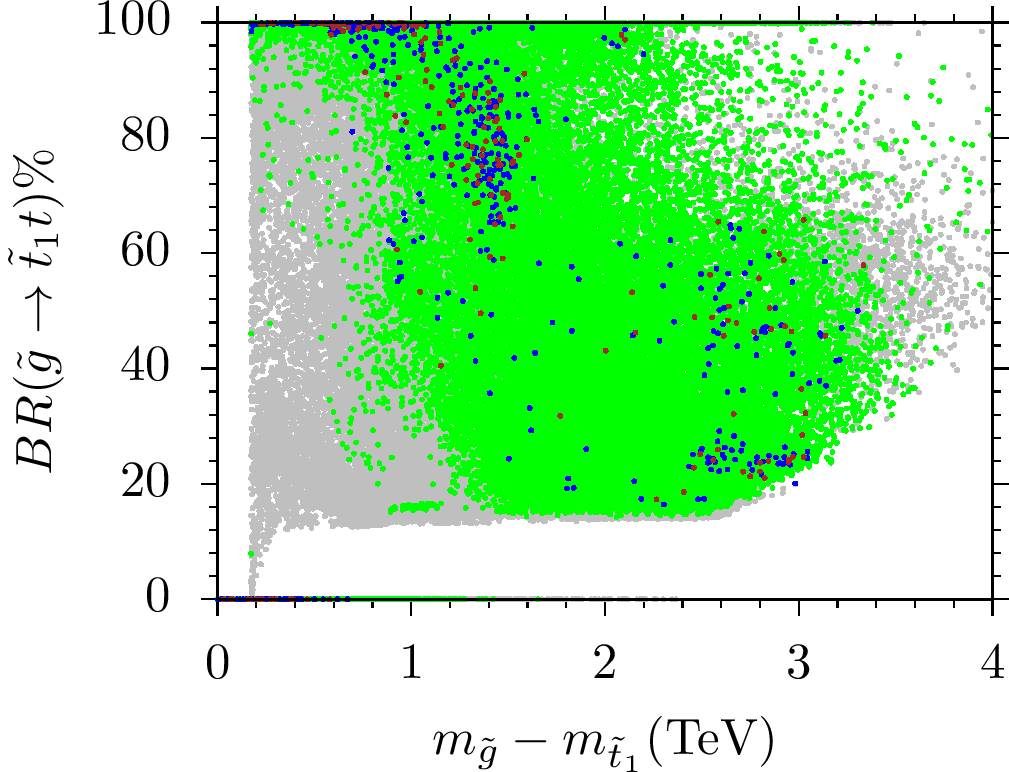}}
\subfigure{\includegraphics[scale=1.1]{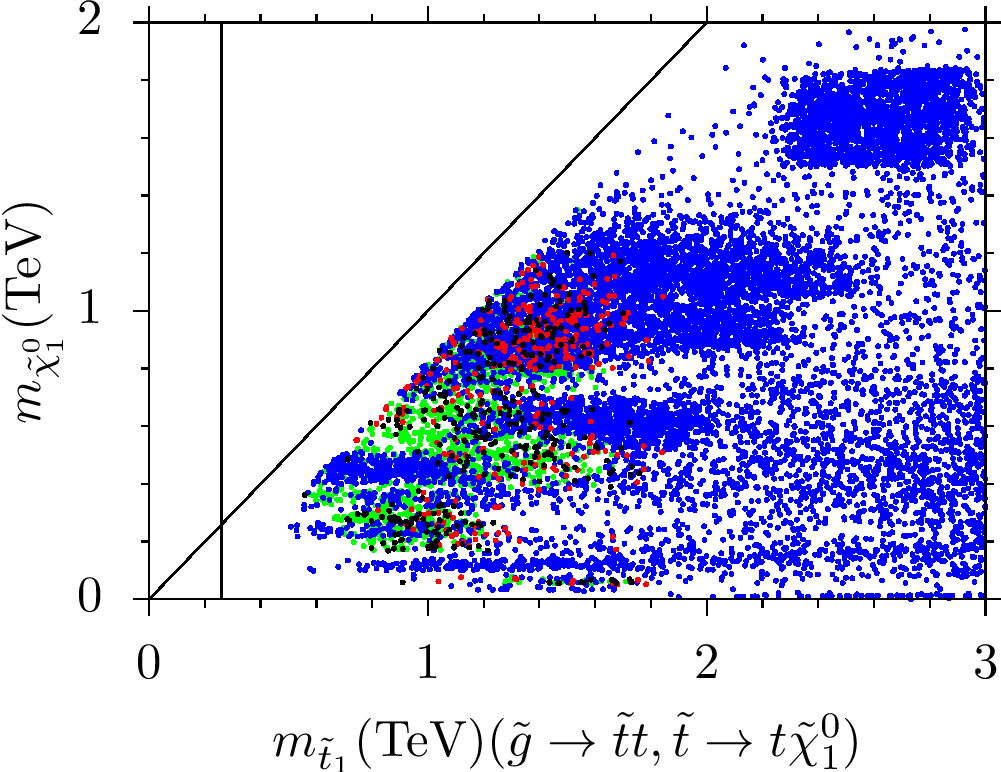}}
\caption{Plots in the ${\rm BR}(\tilde{g}\rightarrow \tilde{t}_{1}t)-{(m_{\tilde{g}}-m_{\tilde{t}_{1}})}$ and $m_{\tilde{\chi}_{1}^{0}}-m_{\tilde{t}_{1}}$ planes. The color coding in the left panel is the same as Figure {in} \ref{fig1}, while the right panel is obtained with the color coding used in Figure \ref{fig5}.}
\label{fig:NUM5}
\end{figure}

In addition to gluino and LSP neutralino, the {decay mode} $\tilde{g}\rightarrow \tilde{t}_{1}t$ is also sensitive to the stop mass, and it leads to {interesting} results {regarding} the stop mass. This is because a SUSY particle exhibits a strong tendency to decay into the {next SUSY} particle in the mass spectrum, if it is allowed. {If} the stop is the heaviest sparticle after gluino in the mass spectrum, the gluino has a strong tendency to decay into a stop as seen from the plots in the ${\rm BR}(\tilde{g}\rightarrow \tilde{t}_{1}t)-{(m_{\tilde{g}}-m_{\tilde{t}_{1}})}$ and ${m_{\tilde{\chi}_{1}^{0}}}-m_{\tilde{t}_{1}}$ planes of Figure \ref{fig:NUM5}. The color coding in the left panel is the same as {in} Figure \ref{fig1}, while the right panel is obtained with the color coding used in Figure \ref{fig5}. The ${\rm BR}(\tilde{g}\rightarrow \tilde{t}_{1}t)-{(m_{\tilde{g}}-m_{\tilde{t}_{1}})}$ plane  shows that ${\rm BR}(\tilde{g}\rightarrow \tilde{t}_{1}t)\sim 100\%$ {for} $m_{\tilde{g}}-m_{\tilde{t}_{1}} \lesssim 1$ TeV, and even though it is possible to obtain {a} large branching fraction for $m_{\tilde{g}}-m_{\tilde{t}_{1}} \sim 2$ TeV, such solutions correspond to {a} heavy gluino, whose pair production becomes {unlikely}. The $m_{\tilde{\chi}_{1}^{0}}-m_{\tilde{t}_{1}}$ {plot} summarizes our discussion for the {decay} $\tilde{g}\rightarrow \tilde{t}_{1}t$ with {the stop subsequently} decaying into a top quark and a LSP neutralino. According to the results shown in this plane, the stop mass {up to about 1.8 TeV} can be {excluded,} if the collider analyses do not yield any direct signal up to this mass scale.

\begin{figure}[ht!]
\subfigure{\includegraphics[scale=1.1]{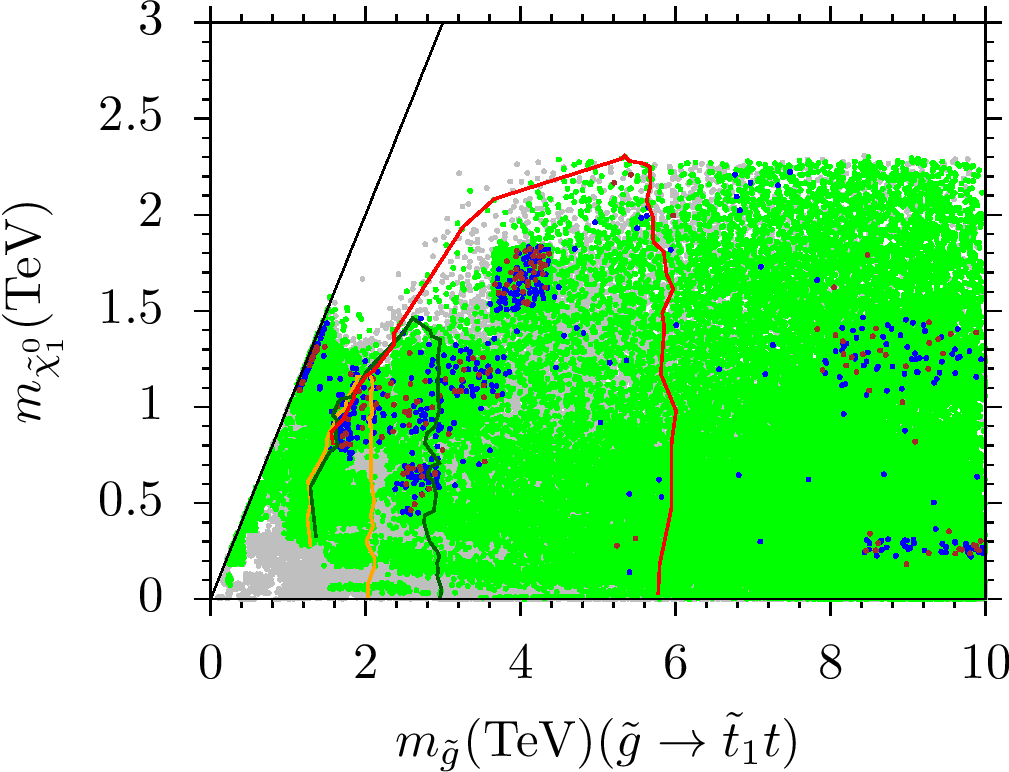}}
\subfigure{\includegraphics[scale=1.1]{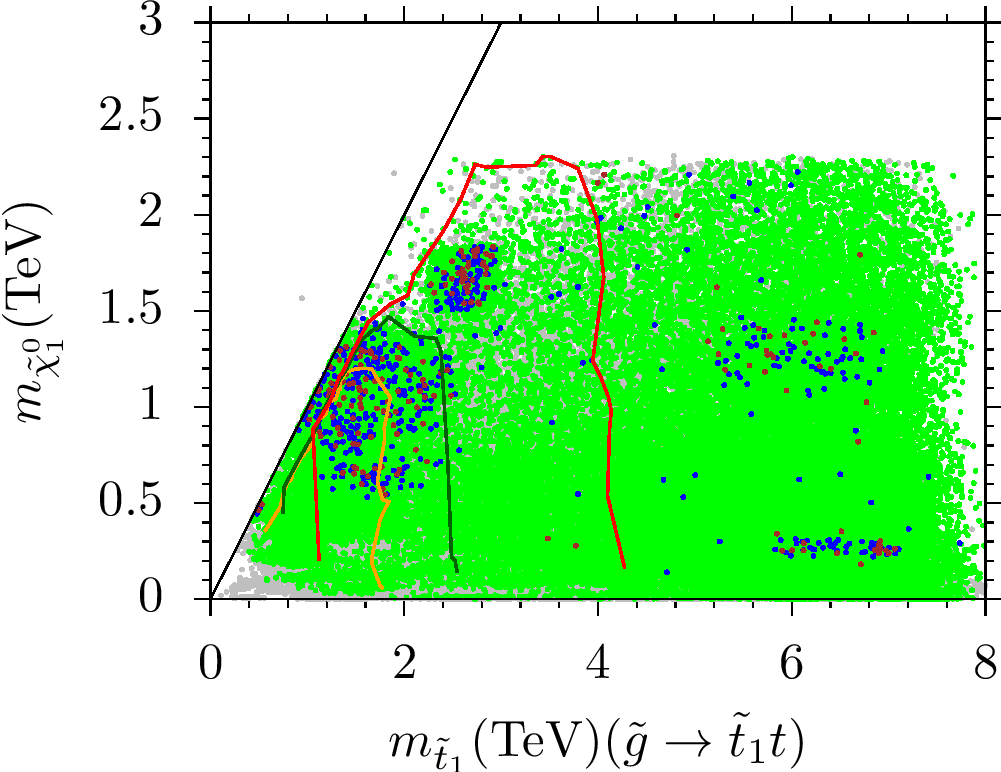}}
\caption{Summary of our {findings} in the $m_{\tilde{\chi}_{1}^{0}}-m_{\tilde{g}}$ and $m_{\tilde{\chi}_{1}^{0}}-m_{\tilde{t}_{1}}$ planes. The color coding is the same as {in} Figure \ref{fig1}. The orange curve represents the exclusion at 14 TeV, while the dark green and the red curves are obtained for 27 TeV and 100 TeV respectively.}
\label{fig:NUGM6}
\end{figure}

Before concluding we summarize our findings in Figure \ref{fig:NUGM6} in the $m_{\tilde{\chi}_{1}^{0}}-m_{\tilde{g}}$ and $m_{\tilde{\chi}_{1}^{0}}-m_{\tilde{t}_{1}}$ planes, {with the} color coding is the same as {in} Figure \ref{fig1}. The orange curve represents the exclusion {region} at 14 TeV, while the dark green and the red curves are obtained for 27 TeV and 100 TeV respectively. The solutions with $m_{\tilde{g}}\lesssim 2$ TeV are excluded in the current LHC run, while the HE-LHC (at 27 TeV) and FCC (at 100 TeV) can probe {a gluino mass} up to about 3 {TeV} and 6 TeV respectively. Comparing with Figure \ref{fig:NUGM_DM} the {exclusion regions} from 14 TeV and 27 TeV do not yield any impact on the stau-neutralino coannihilation and $A-$resonance solutions, while 100 TeV can probe and exclude them in case of no direct signal in near future. Besides, the analyses {for} the $\tilde{g}\rightarrow \tilde{t}_{1}t$ decay mode cannot be applied {if} the gluino happens to be NLSP (brown and blue points around the diagonal line in the left panel). Furthermore, the {decay mode} $\tilde{g}\rightarrow \tilde{t}_{1}t$ can probe the stop up to about 1.8 TeV, 2.5 TeV and 4.4 TeV in the experiments at 14 TeV, 27 TeV and 100 TeV center of mass energies respectively, as shown in the $m_{\tilde{\chi}_{1}^{0}}-m_{\tilde{t}_{1}}$ plane.

\begin{figure}[h!]
\centering
\includegraphics[scale=1.5]{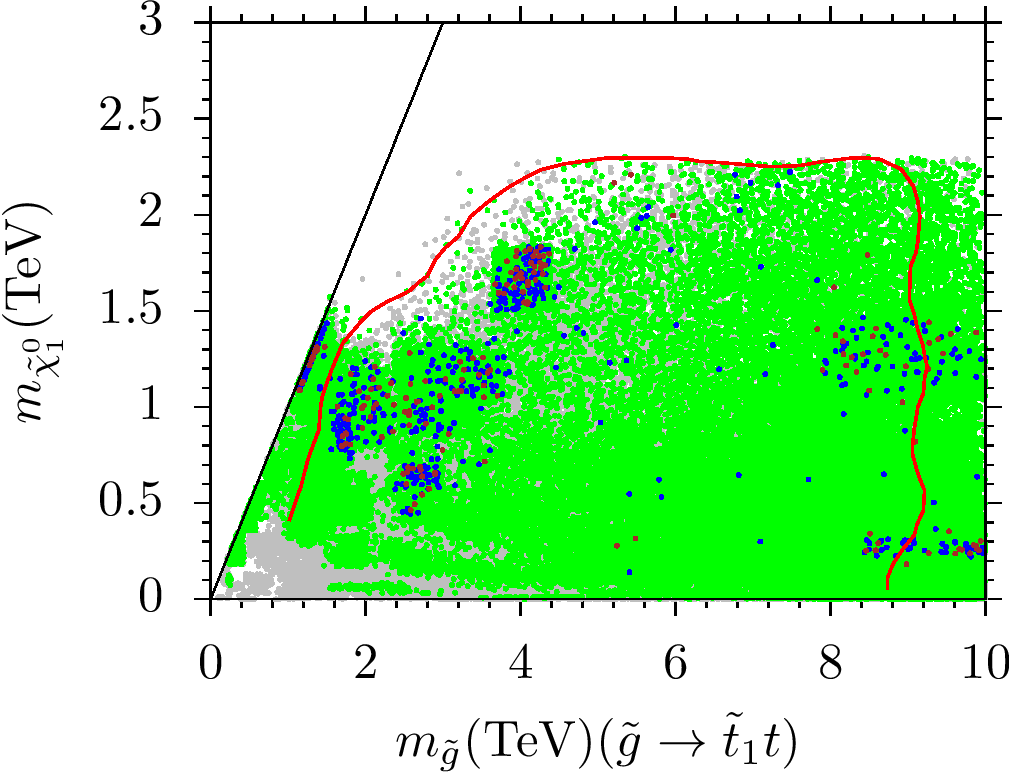}
\caption{Gluino probe at the FCC with 100 TeV center of mass energy. The color coding is the same {in} Figure \ref{fig1}. The red curve represents  the gluino {reach} when the luminosity is set to $3000~fb^{-1}$.}
\label{fig:FCC3000}
\end{figure}

{We should note that the curves represented in Figure \ref{fig:NUGM6} can be interpreted as the minimum reach of the future experiments, since we set the luminosity to $36.1~fb^{-1}$ in our analyses. It has already been reported that the FCC can probe the gluino {mass} up to about 10 TeV (see, for instance, \cite{Baer:2017yqq}) {for an integrated luminosity of} $3000~fb^{-1}$. Figure \ref{fig:FCC3000} displays our results in probing gluino at the FCC experiments {if} the $3000~fb^{-1}$ luminosity is reached. One can expect that {a gluino mass up to about 9-10 TeV} will be {probed,} and we {remain optimistic that a direct signal will sight} the gluino in {the} near future. }

\section{Conclusion}
\label{sec:conc}

{We have explored gluino masses} in the CMSSM and NUGM frameworks through its {decay} into a stop and a top quark, or into a pair of top quarks along with a LSP neutralino. We {find} that the region with $m_{\tilde{g}} \lesssim 2$ TeV  is excluded up to $68\%$ {CL} in the CMSSM {if} the gluino {decays} into a stop and top quark. {The} $95\%$ {CL} exclusion is also quite significant, since it requires $m_{\tilde{g}}\gtrsim 1.9$ TeV. Considering {an} error of about $10\%$ in {the} calculation of the SUSY mass spectrum, such exclusion bounds on the gluino mass more or less overlap with the current LHC results. The {decay mode} $\tilde{g}\rightarrow \bar{t}t\tilde{\chi}_{1}^{0}$ may take over {if} the $\tilde{g}\rightarrow \tilde{t}_{1}t$ is not allowed. One can probe {in this case a} gluino mass up to about 1.5 TeV with $68\%$ CL in the CMSSM, and about 1.4 TeV with $95\%$ CL. {Imposing the DM constraints yield a lower bound on the gluino and stop masses of} about 3.2 TeV, which is beyond the reach of the current LHC experiments.

We {performed a} similar analyses {in the NUGM framework corresponding to non-universal gaugino masses at GUT scale}. Non-universality in the gaugino sector breaks the linear correlation between the gluino and LSP neutralino masses so that one can identify solutions {corresponding to} the gluino-{neutralino} coannihilation scenario. Even though the DM constraints have a strong impact also in the NUGM parameter space, it is not as strong as that in CMSSM, and {a lower bound on the gluino and stop masses is} about 1 TeV. {An NLSP }gluino is realized in the mass {region} $1.1\lesssim m_{\tilde{g}}\approx m_{\tilde{\chi}_{1}^{0}}\lesssim 1.5$ TeV. Since the mass difference between the gluino and LSP neutralino is {rather small}, these solutions cannot be probed through {a gluino decay into a stop and top quark}. Indeed, the exclusion {region in this case from the LHC, namely $m_{\tilde{g}}\gtrsim 800$ GeV,} is not very severe. {In NUGM we identify} stop-neutralino coannihilation solutions with $m_{\tilde{t}_{1}}\approx m_{\tilde{\chi}_{1}^{0}}\in [0.9-1.5]$ TeV {and stau-neutralino coannihilation for $1.5 \lesssim m_{\tilde{\tau}}\lesssim 2$ TeV and $A-$resonance solutions for $2.5 \lesssim m_{A} \lesssim 3$ TeV} . We {find that the a gluino of mass up to about 2 TeV} can be excluded  through the {decay mode} $\tilde{g}\rightarrow \tilde{t}_{1}t$ at 14 TeV, and it can be probed up to about 3 TeV and 6 TeV at the HE-LHC and FCC experiments respectively, {with an integrated luminosity of} 36.1 $fb^{-1}$. We also {note that a 100 TeV FCC can probe gluino masses up to about 9-10 TeV with an integrated luminosity of 3000 $fb^{-1}$}.

Finally, experiments at 100 TeV center of mass energy can also probe the stau-neutralino coannihilation and $A-$resonance solutions. Besides the gluino, the impact on the stop mass through the $\tilde{g}\rightarrow \tilde{t}_{1}t$ channel is significant as it can be probed up to about 1.8 TeV, 2.5 TeV and 4.4 TeV {in experiments with} 14 TeV, 27 TeV and 100 TeV center of mass energies respectively.

\section{Acknowledgments}

The work of Z. A, Z. K. and C. S. U is supported by the Scientific and Technological Research Council of Turkey (TUBITAK) Grant no. MFAG-118F090. {Q. S. is supported in part} by DOE under Grant no. DE-SC 0013880. CSU also would like to thank the Physics and Astronomy Department and Bartol Research Institute of the University of Delaware, for kind hospitality where part of this work has been done. Part of the calculations reported in this paper were performed at the National Academic Network and Information Center (ULAKBIM) of TUBITAK, High Performance and Grid Computing Center (TRUBA Resources).


\begin{thebibliography}{99}

\bibitem{big-422}
B. Ananthanarayan, G. Lazarides and Q. Shafi, Phys. Rev. D {\bf 44},
1613 (1991) and Phys. Lett. B {\bf 300}, 24 (1993)5; Q.~Shafi and
B.~Ananthanarayan, Trieste HEP Cosmol.1991:233-244;

\bibitem{bigger-422}
V. Barger, M. Berger and P. Ohmann, Phys. Rev. D {\bf 49}, (1994)
4908; M. Carena, M. Olechowski, S. Pokorski and C. Wagner,  Nucl.\
Phys.\  B {\bf 426}, 269 (1994); B. Ananthanarayan, Q. Shafi and X.
Wang, Phys. Rev. D {\bf 50}, 5980 (1994); G. Anderson et al. Phys.
Rev. D {\bf 47}, (1993) 3702 and Phys. Rev. D {\bf 49},  3660
(1994); R. Rattazzi and U. Sarid, Phys. Rev. D {\bf 53}, 1553
(1996); T. Blazek, M. Carena, S. Raby and C. Wagner, Phys. Rev. D
{\bf 56}, 6919 (1997); T. Blazek, S. Raby and K. Tobe, Phys. Rev. D
{\bf 62}, 055001 (2000); H. Baer, M. Diaz, J. Ferrandis and X. Tata,
Phys. Rev. D {\bf 61}, 111701 (2000); H. Baer, M. Brhlik, M. Diaz,
J. Ferrandis, P. Mercadante, P. Quintana and X. Tata, Phys. Rev. D
{\bf 63}, 015007(2001); S. Profumo, Phys. Rev. D {\bf 68} (2003)
015006; C.~Balazs and R.~Dermisek, JHEP {\bf 0306}, 024 (2003);
C. Pallis, Nucl. Phys. B {\bf 678},  398 (2004); M. Gomez,
G. Lazarides and C. Pallis, Phys. Rev. D {\bf 61} (2000) 123512,
Nucl. Phys. B {\bf 638},  165 (2002) and Phys. Rev. D {\bf 67},
097701(2003);  I.~Gogoladze, Y.~Mimura, S.~Nandi and K.~Tobe, Phys.\
Lett.\  B {\bf 575}, 66 (2003); U. Chattopadhyay, A. Corsetti and P.
Nath, Phys. Rev. D {\bf 66} 035003, (2002); T.~Blazek, R.~Dermisek
and S.~Raby, Phys.\ Rev.\ Lett.\  {\bf 88}, 111804 (2002) and Phys.\
Rev.\  D {\bf 65}, 115004 (2002); M. Gomez, T. Ibrahim, P. Nath and
S. Skadhauge, Phys. Rev. D {\bf 72}, 095008 (2005); K. Tobe and J.
D. Wells, Nucl. Phys. B {\bf 663}, 123 (2003); W.~Altmannshofer,
D.~Guadagnoli, S.~Raby and D.~M.~Straub, Phys.\ Lett.\  B {\bf 668},
385 (2008); D.~Guadagnoli, S.~Raby and D.~M.~Straub, JHEP {\bf 0910}, 059 (2009);
 H.~Baer, S.~Kraml and S.~Sekmen, JHEP {\bf 0909}, 005 (2009);
  K.~Choi, D.~Guadagnoli, S.~H.~Im and C.~B.~Park,
  arXiv:1005.0618 [hep-ph];B.~Dutta and Y.~Mimura,
  arXiv:1810.08413 [hep-ph];
  S.~Raza, Q.~Shafi and C.~S.~Ün,
  Phys.\ Rev.\ D {\bf 92}, no. 5, 055010 (2015)
  [arXiv:1412.7672 [hep-ph]].
\bibitem{Chattopadhyay:2001mj}
 U.~Chattopadhyay and P.~Nath,
  Phys.\ Rev.\  D {\bf 65}, 075009 (2002);
  S.~Komine and M.~Yamaguchi,
  Phys.\ Rev.\  D {\bf 65}, 075013 (2002);
  S.~Profumo,
  Phys.\ Rev.\  D {\bf 68}, 015006 (2003);
  C.~Pallis,
  Nucl.\ Phys.\  B {\bf 678}, 398 (2004);
  C.~Balazs and R.~Dermisek,
  JHEP {\bf 0306}, 024 (2003);
  W.~Altmannshofer, D.~Guadagnoli, S.~Raby and D.~M.~Straub,
  Phys.\ Lett.\  B {\bf 668}, 385 (2008);
  I.~Gogoladze, R.~Khalid, N.~Okada and Q.~Shafi,
  Phys.\ Rev.\  D {\bf 79}, 095022 (2009);
  S.~Antusch and M.~Spinrath,
  Phys.\ Rev.\  D {\bf 79}, 095004 (2009);
  H.~Baer, I.~Gogoladze, A.~Mustafayev, S.~Raza and Q.~Shafi,
  JHEP {\bf 1203}, 047 (2012)
  [arXiv:1201.4412 [hep-ph]]; 
  I.~Gogoladze, R.~Khalid, S.~Raza and Q.~Shafi,
  JHEP {\bf 1012}, 055 (2010)
  [arXiv:1008.2765 [hep-ph]];
  S.~Raza, Q.~Shafi and C.~S.~Un,
  JHEP {\bf 1905}, 046 (2019)
  [arXiv:1812.10128 [hep-ph]].

\bibitem{Langacker:1998tc} 
  P.~Langacker and J.~Wang,
  Phys.\ Rev.\ D {\bf 58}, 115010 (1998)
  [hep-ph/9804428];
  S.~M.~Barr,
  Phys.\ Rev.\ Lett.\  {\bf 55}, 2778 (1985);
  doi:10.1103/PhysRevLett.55.2778
  J.~L.~Hewett and T.~G.~Rizzo,
  Phys.\ Rept.\  {\bf 183} (1989) 193;
  doi:10.1016/0370-1573(89)90071-9
  M.~Cvetic and P.~Langacker,
  Phys.\ Rev.\ D {\bf 54}, 3570 (1996)
  doi:10.1103/PhysRevD.54.3570
  [hep-ph/9511378];
  G.~Cleaver, M.~Cvetic, J.~R.~Espinosa, L.~L.~Everett and P.~Langacker,
  Phys.\ Rev.\ D {\bf 57}, 2701 (1998)
  doi:10.1103/PhysRevD.57.2701
  [hep-ph/9705391];
  G.~Cleaver, M.~Cvetic, J.~R.~Espinosa, L.~L.~Everett and P.~Langacker,
  Nucl.\ Phys.\ B {\bf 525}, 3 (1998)
  doi:10.1016/S0550-3213(98)00277-6
  [hep-th/9711178];
  D.~M.~Ghilencea, L.~E.~Ibanez, N.~Irges and F.~Quevedo,
  JHEP {\bf 0208}, 016 (2002)
  doi:10.1088/1126-6708/2002/08/016
  [hep-ph/0205083];
  S.~F.~King, S.~Moretti and R.~Nevzorov,
  Phys.\ Rev.\ D {\bf 73}, 035009 (2006)
  doi:10.1103/PhysRevD.73.035009
  [hep-ph/0510419];
  R.~Diener, S.~Godfrey and T.~A.~W.~Martin,
  arXiv:0910.1334 [hep-ph];
  P.~Langacker,
  Rev.\ Mod.\ Phys.\  {\bf 81}, 1199 (2009)
  doi:10.1103/RevModPhys.81.1199
  [arXiv:0801.1345 [hep-ph]];

\bibitem{Aaboud:2017vwy} 
  M.~Aaboud {\it et al.} [ATLAS Collaboration],
  Phys.\ Rev.\ D {\bf 97}, no. 11, 112001 (2018)
  doi:10.1103/PhysRevD.97.112001
  [arXiv:1712.02332 [hep-ex]];
  M.~Aaboud {\it et al.} [ATLAS Collaboration],
  Phys.\ Rev.\ D {\bf 96}, no. 11, 112010 (2017)
  doi:10.1103/PhysRevD.96.112010
  [arXiv:1708.08232 [hep-ex]].


\bibitem{Vami:2019slp} 
  T.~A.~Vami [ATLAS and CMS Collaborations],
  arXiv:1909.11753 [hep-ex].
  
\bibitem{Cici:2016oqr} 
  A.~\c{C}i\c{c}i, Z.~Kırca and C.~S.~\"{U}n,
  Eur.\ Phys.\ J.\ C {\bf 78}, no. 1, 60 (2018)
  doi:10.1140/epjc/s10052-018-5549-y
  [arXiv:1611.05270 [hep-ph]].



\bibitem{Leblanc:2018rfd} 
  M.~LeBlanc [ATLAS Collaboration],
  PoS DIS {\bf 2018}, 078 (2018);
  M.~Aaboud {\it et al.} [ATLAS Collaboration],
  Phys.\ Rev.\ D {\bf 99}, no. 1, 012009 (2019)
  [arXiv:1808.06358 [hep-ex]];
  M.~Aaboud {\it et al.} [ATLAS Collaboration],
  Phys.\ Rev.\ D {\bf 97}, no. 11, 112001 (2018)
  [arXiv:1712.02332 [hep-ex]].



\bibitem{Aad:2016eki} 
  G.~Aad {\it et al.} [ATLAS Collaboration],
  Phys.\ Rev.\ D {\bf 94}, no. 3, 032003 (2016)
  doi:10.1103/PhysRevD.94.032003
  [arXiv:1605.09318 [hep-ex]].

\bibitem{TheATLAScollaboration:2013aia} 
  The ATLAS collaboration [ATLAS Collaboration],
  ATLAS-CONF-2013-068.

\bibitem{Porod:2003um}
  Porod, W.
  {\it Comput.\ Phys.\ Commun.\ } {\bf 2003}, {\it 153}, 275.

\bibitem{Porod2}
  Porod, W. and Staub, F.
  {\it Comput.\ Phys.\ Commun.\ } {\bf 2012} {\it 183}, 2458.

\bibitem{Staub:2008uz}
  Staub, F.
{\bf 2008}, {\it Preprint  arXiv:0806.0538}.
\bibitem{Staub2}
  Staub, F.
  {\it Comput.\ Phys.\ Commun.\ } {\bf 2011} {\it 182}, 808.

\bibitem{Hisano:1992jj}
  Hisano, J.; Murayama, H.; and Yanagida, T.
 {\it Nucl.\ Phys.\ B } {\bf 1993} {\it 402}, 46.

\bibitem{GUTth}
  Chkareuli, J. L.; and Gogoladze, I. G.
 {\it Phys.\ Rev.\ D }{\bf 1998} {\it 58}, 055011.

\bibitem{Group:2009ad}
  T.~E.~W.~Group [CDF and D0 Collaborations], {\bf 2009}, {\it Preprint
  arXiv:0903.2503}.

\bibitem{Gogoladze:2011db}
  Gogoladze, I.; Khalid, R.; Raza S.; and Shafi Q.
  {\it JHEP } {\bf 2011}, {\it 1106}, 117.

\bibitem{Gogoladze:2011aa}
  Gogoladze, I.; Shafi, Q.; and Un, C. S.
 {\it JHEP} {\bf 2012} {\it 1208}, 028.

\bibitem{Ajaib:2013zha}
  Adeel Ajaib, M.; Gogoladze, I.; Shafi, Q.; and Un, C. S.
  {\it JHEP } {\bf 2013} {\it 1307}, 139.

\bibitem{Ibanez:Ross}
  Ibanez, L. E.; and Ross, G. G.
  {\it Phys.\ Lett.} {\it 110B} {\bf 1982} 215.

\bibitem{REWSB2}
  Inoue, K.; Kakuto, A.; Komatsu, H.; and Takeshita S.,
  {\it Prog.\ Theor.\ Phys.\ }  {\bf 1982} {\it 68}, 927.

\bibitem{REWSB3}
  Ibanez, L. E.
  {\it Phys.\ Lett.\ } {\bf 1982} {\it 118B}, 73.
  
\bibitem{REWSB4}  
  Ellis, J. R.; Nanopoulos D. V.; and Tamvakis, K.
 {\it Phys.\ Lett.\ } {\bf 1983} {\it 121B}, 123.
 
 \bibitem{REWSB5}
   Alvarez-Gaume, L.; Polchinski, J.; and Wise, M. B.
 {\it Nucl.\ Phys.\ B} {\bf 1983} {\it 221}, 495.

\bibitem{Nakamura:2010zzi}
  Nakamura, K. {\it et al.} [Particle Data Group Collaboration],
  {\it J.\ Phys.\ G} {\bf 2010} {\it 37}, 075021.

\bibitem{Hinshaw:2012aka} 
  G.~Hinshaw {\it et al.} [WMAP Collaboration],
  Astrophys.\ J.\ Suppl.\  {\bf 208}, 19 (2013)
  [arXiv:1212.5226 [astro-ph.CO]].


\bibitem{Akrami:2018vks}
  Y.~Akrami {\it et al.} [Planck Collaboration],
  arXiv:1807.06205 [astro-ph.CO].

\bibitem{Baer:2012by} See for instance;

  H.~Baer, I.~Gogoladze, A.~Mustafayev, S.~Raza and Q.~Shafi,
  JHEP {\bf 1203}, 047 (2012)
  doi:10.1007/JHEP03(2012)047
  [arXiv:1201.4412 [hep-ph]];
  T.~Li, D.~V.~Nanopoulos, S.~Raza and X.~C.~Wang,
  JHEP {\bf 1408}, 128 (2014)
  doi:10.1007/JHEP08(2014)128
  [arXiv:1406.5574 [hep-ph]].


\bibitem{Belanger:2006is} 
  G.~Belanger, F.~Boudjema, A.~Pukhov and A.~Semenov,
  Comput.\ Phys.\ Commun.\  {\bf 176}, 367 (2007)
  doi:10.1016/j.cpc.2006.11.008
  [hep-ph/0607059];
  G.~Belanger, F.~Boudjema, A.~Pukhov and A.~Semenov,
  Comput.\ Phys.\ Commun.\  {\bf 185}, 960 (2014)
  doi:10.1016/j.cpc.2013.10.016
  [arXiv:1305.0237 [hep-ph]].

\bibitem{Alwall:2011uj} 
  J.~Alwall, M.~Herquet, F.~Maltoni, O.~Mattelaer and T.~Stelzer,
  JHEP {\bf 1106}, 128 (2011)
  doi:10.1007/JHEP06(2011)128
  [arXiv:1106.0522 [hep-ph]].

\bibitem{Agashe:2014kda}
  Olive, K. A. {\it et al.} [Particle Data Group Collaboration],
  {\it Chin.\ Phys.\ C } {\bf 2014} {\it 38}, 090001.

\bibitem{Aaij:2012nna}
  Aaij, R. {\it et al.} [LHCb Collaboration],
  {\it Phys.\ Rev.\ Lett.\ } {\bf 2013}  {\it 110}, no. 2, 021801.

\bibitem{Amhis:2012bh}
  Amhis, Y. {\it et al.} [Heavy Flavor Averaging Group Collaboration],
 {\bf 2012}, {\it Preprint arXiv:1207.1158}.
\bibitem{Asner:2010qj}
  Asner, D. {\it et al.} [Heavy Flavor Averaging Group Collaboration],
{\bf 2010 }, {\it Preprint arXiv:1010.1589}.

\bibitem{Bisset:1996sb} 
  M.~Bisset, S.~Raychaudhuri and D.~K.~Ghosh,
  hep-ph/9608421.

\bibitem{Cranmer:2015nia} 
  K.~Cranmer,
  doi:10.5170/CERN-2015-001.247, 10.5170/CERN-2014-003.267
  arXiv:1503.07622 [physics.data-an].





\bibitem{Belanger:2009ti}
  Belanger, G.; Boudjema, F.; Pukhov, A.; and Singh, R. K.
  {\it JHEP } {\bf 2009} {\it 0911}, 026.
  
  \bibitem{SekmenMH}
  Baer, H.; Kraml, S.; Sekmen, S.; and Summy, H.
  {\it JHEP } {\bf 2008} {\it 0803}, 056.






\bibitem{Baer:2016wkz} See, for instance,\\
  H.~Baer, V.~Barger, J.~S.~Gainer, P.~Huang, M.~Savoy, D.~Sengupta and X.~Tata,
  Eur.\ Phys.\ J.\ C {\bf 77}, no. 7, 499 (2017)
  doi:10.1140/epjc/s10052-017-5067-3
  [arXiv:1612.00795 [hep-ph]].

\bibitem{Borschensky:2014cia} 
  C.~Borschensky, M.~Krämer, A.~Kulesza, M.~Mangano, S.~Padhi, T.~Plehn and X.~Portell,
  Eur.\ Phys.\ J.\ C {\bf 74}, no. 12, 3174 (2014)
  doi:10.1140/epjc/s10052-014-3174-y
  [arXiv:1407.5066 [hep-ph]] \\ (see also \href{}{https://twiki.cern.ch/twiki/bin/view/LHCPhysics/SUSYCrossSections})

\bibitem{Baer:2017yqq} 
  H.~Baer, V.~Barger, J.~S.~Gainer, P.~Huang, M.~Savoy, H.~Serce and X.~Tata,
  Phys.\ Lett.\ B {\bf 774}, 451 (2017)
  [arXiv:1702.06588 [hep-ph]];
  H.~Baer, V.~Barger, J.~S.~Gainer, H.~Serce and X.~Tata,
  Phys.\ Rev.\ D {\bf 96}, no. 11, 115008 (2017)
  [arXiv:1708.09054 [hep-ph]].





\bibitem{Martin:2009ad}
 B.~Ananthanarayan, P.~N.~Pandita,
  Int.\ J.\ Mod.\ Phys.\  {\bf A22}, 3229-3259 (2007);
  S.~Bhattacharya, A.~Datta and B.~Mukhopadhyaya,
  JHEP {\bf 0710}, 080 (2007);
   S.~P.~Martin,
  Phys.\ Rev.\  {\bf D79}, 095019 (2009);
  J.~Chakrabortty and A.~Raychaudhuri,
  Phys.\ Lett.\ B {\bf 673}, 57 (2009).

\bibitem{Martin:2013aha}
  S.~P.~Martin,
  arXiv:1312.0582 [hep-ph].

\bibitem{Anandakrishnan:2013cwa}
  A.~Anandakrishnan and S.~Raby,
  Phys.\ Rev.\ Lett.\  {\bf 111}, 211801 (2013);
  S.~Raza, Q.~Shafi and C.~S.~Un,
  arXiv:1812.10128 [hep-ph], and references therein.




\end{thebibliography}
\end{document}